%% file: iclr2026_conference.tex
\documentclass{article} % For LaTeX2e
\usepackage{iclr2026_conference,times}

\input{math_commands.tex}

\usepackage{hyperref}
\usepackage{xurl}
\usepackage{booktabs}
\usepackage{wrapfig}
\usepackage{graphicx}
\usepackage{subcaption}
\usepackage{xcolor}
\usepackage{multirow}
\usepackage[export]{adjustbox}
\usepackage{amssymb}
\usepackage{graphicx}
\usepackage{caption}
\usepackage{makecell}
\usepackage{tabularx}
\usepackage{array}
\usepackage{nicefrac}
\usepackage{tcolorbox}
\usepackage{placeins}
\newcommand{\headcell}[1]{%
  \begin{tabular}[c]{@{}c@{}}#1\end{tabular}%
}

\title{Machine Hearing: Can Audio LLMs Generalize Beyond Human Auditory Perception?}

\author{%
\textbf{Liwei Liu}$^{1,2}$, \textbf{Jun Tao}$^{1,3}$,
\textbf{Xingyuan Li}$^{1,4}$, \textbf{Xuenan Xu}$^{1}$,
\textbf{Chen Zhang}$^{1,5}$, \textbf{Junyi Wang}$^{1,6}$, \\
\textbf{Boya Zhang}$^{1,6}$, \textbf{Xiaoyu Yang}$^{7}$,
\textbf{Zhen-Hua Ling}$^{3}$, \textbf{Wen Wu}$^{1*}$,
\textbf{Chao Zhang}$^{1,5*}$ \\[0.8em]
{\small $^{1}$Shanghai Artificial Intelligence Laboratory} \\
{\small $^{2}$Harbin Institute of Technology} \\
{\small $^{3}$University of Science and Technology of China} \\
{\small $^{4}$Fudan University} \\
{\small $^{5}$Tsinghua University} \\
{\small $^{6}$Shanghai Jiao Tong University} \\
{\small $^{7}$University of Cambridge} \\[0.5em]
{\small \texttt{\{liuliwei,wuwen\}@pjlab.org.cn}} \\
{\small $^{*}$Corresponding authors}
}

\usepackage{xcolor}

\newif\ifshowcommentmarks
\showcommentmarksfalse % 投稿前关闭

\newcommand{\cmt}{%
  \ifshowcommentmarks
    \textcolor{magenta}{\textbf{[COMMENT]}}%
  \fi
}

\iclrfinalcopy % Uncomment for camera-ready version, but NOT for submission.
\begin{document}

\maketitle
\lhead{Preprint}
\begin{abstract}
Multimodal LLMs have made remarkable progress in perceiving the physical world, while their perceptual scope remains largely human-centric, focusing on what humans can directly see or hear. This progress raises the prospect of perception beyond human limits, such as extending hearing to sounds that are inaudible to humans. More broadly, the physical world is full of waves, vibrations, and oscillations, only a small fraction of which are directly audible to humans. This invites the extension of multimodal LLMs toward a broader form of machine hearing, which remains largely underexplored. To investigate this direction, we introduce the \textbf{M}achine \textbf{H}earing \textbf{B}enchmark (\textbf{MHB}\footnote{The benchmark is available at \url{https://huggingface.co/datasets/OpenTSLab/MachineHearing}.}), comprising 6,128 instances across 13 tasks and seven domains. 
% MHB spans highly heterogeneous input regimes, with sampling rates from approximately \(10^{-5}\) to \(10^{6}\) Hz, sequence lengths spanning approximately \(10^{1}\) to \(10^{6}\) samples, and both single- and multi-channel inputs. 
MHB spans highly heterogeneous input regimes, with sampling rates from approximately \(10^{-5}\) to \(10^{6}\) Hz, sequence lengths from \(10^{1}\) to \(10^{6}\) samples, and both single- and multi-channel inputs. 
Evaluations of current open-source and proprietary Audio LLMs (ALLMs) reveal limited zero-shot transfer to machine hearing. Supervised adaptation on machine-hearing data alone does not close this gap, suggesting that the challenge extends beyond data exposure to a systematic mismatch between existing audio front ends and machine-hearing inputs. To address this mismatch, we introduce \textbf{MachEar}, a lightweight encoder extension that complements the pretrained Audio LLM for machine-hearing inputs. The results show that MachEar equips a pretrained Audio LLM with stronger machine-hearing capabilities while largely preserving its existing audio understanding performance.

\end{abstract}

\section{Introduction}
\label{sec:intro}
\begin{figure}[t]
  \centering
  \vspace{-2.4em}
  \includegraphics[width=\linewidth]{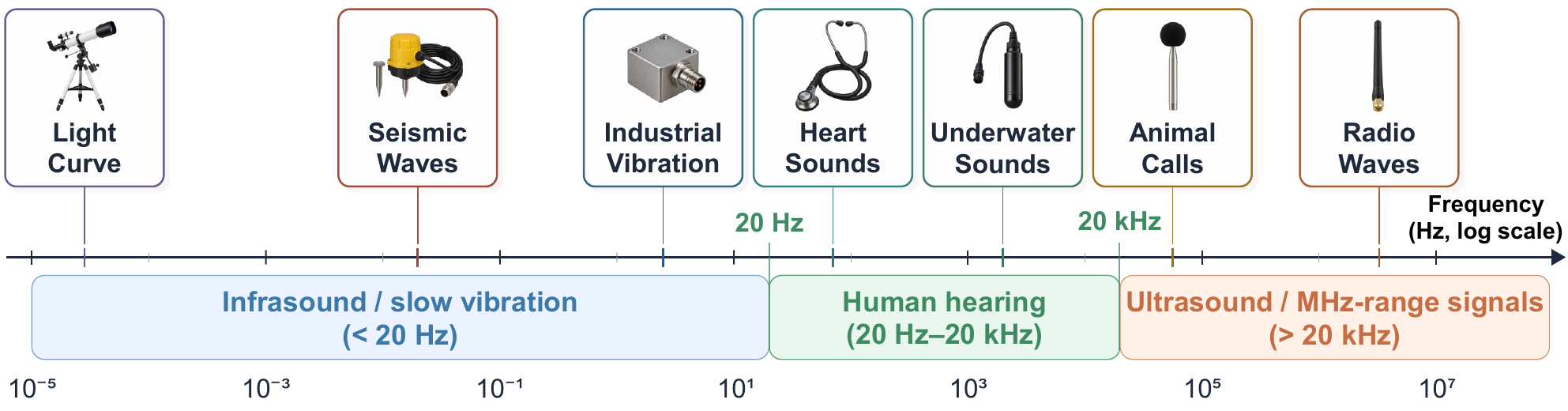}
  \caption{
  Conceptual illustration of waves, vibrations, and oscillations across characteristic frequency scales. Human-audible sound constitutes only a small subset of the broader signal space considered in machine hearing.\cmt
  }
  \label{fig:mhb-frequency}
  \vspace{-1.0em}
\end{figure}

% llw
Advanced Audio Large Language Models (ALLMs)~\citep{chu2024qwen2audio,tang2024salmonn,xu2025qwen25omni} have demonstrated strong capabilities in understanding speech, music, environmental sounds, and other forms of audio. However, their perceptual scope remains largely human-centric, focusing primarily on what humans themselves can hear. As multimodal intelligence continues to expand, a natural question is whether this boundary should remain tied to human auditory perception.
The physical world is full of waves, vibrations, and oscillations, only a small fraction of which are directly audible to humans (Fig.~\ref{fig:mhb-frequency}). Earthquakes rumble through the ground, machines vibrate as they operate, radio and radar waves carry oscillatory patterns, biological systems pulse and resonate, and even astronomical phenomena vary in characteristic rhythms. Extending ALLMs to interpret such phenomena therefore opens a broader form of machine hearing, raising a fundamental question: can current ALLMs extend beyond human-audible audio to understand what humans cannot hear?

To systematically investigate this question, we introduce the \textbf{M}achine \textbf{H}earing \textbf{B}enchmark (\textbf{MHB}), comprising 6,128 instances across 13 tasks and seven domains: Astronomy, Earth Science, Healthcare, Electromagnetic, Underwater Acoustics, Bioacoustics, and Manufacturing. 
% As illustrated in Fig.~\ref{fig:mhb-frequency}, 
% MHB covers a wide range of machine-hearing phenomena, spanning sampling rates from approximately \(10^{-5}\) to \(10^{6}\) Hz, sequence lengths spanning approximately \(10^{1}\) to \(10^{6}\) samples, and both single- and multi-channel inputs. 
MHB covers a wide range of machine-hearing phenomena, with sampling rates from approximately \(10^{-5}\) to \(10^{6}\) Hz, sequence lengths from \(10^{1}\) to \(10^{6}\) samples, and both single- and multi-channel inputs. 
This diversity allows us to examine whether current ALLMs can extend beyond conventional human-audible audio across fundamentally different input regimes. 

We benchmark seven current ALLMs, including both open-source and proprietary models, on MHB to examine how far their existing hearing capabilities extend beyond conventional audio. The results reveal a striking gap. Even taking the best-performing Audio LLM for each task, performance remains below random guessing on 9 of the 13 tasks. Moreover, the difficulty is not limited to incorrect predictions: existing models can fail to recognize or process machine-hearing inputs, with such failures particularly concentrated in regimes with extreme sampling rates or unusually short sequences. We further ask whether these limitations can be addressed simply by exposing ALLMs to machine-hearing data. Supervised adaptation substantially improves performance, but does not consistently eliminate the gap. The remaining failures instead point to systematic limitations in how ALLMs process inputs far outside conventional audio regimes.

% llw

% To address these gaps, we introduce the Machine Hearing Benchmark (MHB), the first unified benchmark for evaluating Machine Hearing across heterogeneous sensor-recorded signals. MHB comprises nearly XXK samples from XX representative tasks covering seven sensing domains, including astronomy, geophysics, underwater acoustics, bioacoustics, industrial monitoring, biomedical sensing, and electromagnetic signal analysis. Building upon MHB, we further propose an enhanced Audio LLM framework that augments existing audio encoders with lightweight signal-aware branches, enabling more effective understanding of heterogeneous physical signals while preserving their original capability on conventional audio tasks.

% llw
Motivated by these findings, we introduce \textbf{MachEar}, a lightweight encoder extension designed to complement the pretrained audio encoder in low- and high-frequency regimes and on extremely short sequences. Equipping Qwen2.5-Omni with MachEar improves its weighted Macro-F1 on MHB from 67.02\% to 72.30\%\cmt, while largely preserving its ASR and audio captioning performance.

\textbf{Summary of Contributions.}
\textbf{1)} We construct the Machine Hearing Benchmark (\textbf{MHB}), which extends the evaluation of ALLMs beyond conventional human-audible audio to a broad range of machine-hearing phenomena across diverse domains and input regimes.
\textbf{2)} We systematically benchmark current open-source and proprietary Audio LLMs on MHB, revealing limited transfer beyond human-audible audio and showing that supervised adaptation alone is insufficient to overcome these limitations.
\textbf{3)} We introduce \textbf{MachEar}, a lightweight encoder extension that equips pretrained Audio LLMs with stronger machine-hearing capabilities while largely preserving their existing audio understanding performance.

% ============================================================
% Figure 1: Frequency coverage
% ============================================================

% llw

% Using MHB, we systematically evaluate seven state-of-the-art audio-capable LLMs together with five representative foundation models. Our results show that current Audio LLMs remain limited in Machine Hearing, whereas domain-specific foundation models provide complementary signal understanding capabilities. We therefore augment existing Audio LLMs with signal-aware architectural adaptations to better accommodate heterogeneous sensor-recorded signals. The proposed framework achieves state-of-the-art performance on MHB while maintaining its performance on conventional speech and audio understanding tasks.

\raggedbottom
\section{Related Work \cmt}
% Recent years have witnessed rapid progress in both benchmark construction and foundation model development for audio and sensor signal understanding. Existing efforts have substantially advanced machine perception across speech, acoustics, and a variety of domain-specific sensing applications. In this section, we review prior work from two complementary perspectives: benchmarks for evaluating machine hearing capabilities and foundation models for understanding auditory and sensor signals. Together, these studies provide the foundation upon which we build a unified machine audition benchmark and a generalized learning framework.

% llw
\subsection{Audio Benchmarks} 
Existing audio benchmarks, such as SUPERB~\citep{yang2021superb}, SUPERB-SG~\citep{tsai2022superb}, HEAR~\citep{turian2022hear}, HARES~\citep{wang2022towards}, AIR-Bench~\citep{yang2024air}, and AudioBench~\citep{wang2025audiobench}, primarily evaluate speech, music, environmental sounds, and other forms of human-audible audio. \cmt
Within this scope, these benchmarks evaluate models on downstream tasks such as automatic speech recognition (ASR), audio captioning, sound event recognition, and speech emotion recognition. 
As a result, current audio evaluation remains largely human-centric, with limited coverage of phenomena and input regimes beyond conventional human hearing. 
Whether Audio LLMs can extend beyond this boundary toward a broader form of machine hearing therefore remains largely underexplored. MHB complements existing audio benchmarks by extending evaluation beyond human-audible audio to diverse machine-hearing phenomena across a wide range of domains and input regimes, within a unified LLM-compatible framework.
% llw
% Audio benchmarks, such as SUPERB~\citep{yang2021superb}, HEAR~\citep{turian2022hear}, and AudioBench~\citep{wang2025audiobench}, have established standardized evaluation protocols for speech, music, environmental sounds, and other human hearing tasks, greatly facilitating the development and comparison of audio foundation models and Audio LLMs. Beyond conventional audio, numerous benchmarks have been introduced for specialized sensing applications, including industrial monitoring, biomedical signal analysis, geophysical observation, radio-frequency sensing, and astronomical signal analysis. These benchmarks have significantly advanced research within their respective domains by providing task-specific datasets and evaluation protocols tailored to individual sensing modalities. While existing benchmarks cover a broad spectrum of sensing applications, they are typically developed independently for specific domains, making cross-domain evaluation difficult. Consequently, the generalization ability of foundation models across heterogeneous sensor signals has not yet been systematically investigated. Our Machine Hearing Benchmark complements existing efforts by providing a unified evaluation framework spanning diverse physical-world sensing modalities, enabling comprehensive assessment of generalized machine audition.

\subsection{Audio LLMs} 
Audio LLMs, such as Qwen2-Audio~\citep{chu2024qwen2audio}, SALMONN~\citep{tang2024salmonn}, Audio Flamingo 3~\citep{ghosh2025audioflamingo3}, and Qwen2.5-Omni~\citep{xu2025qwen25omni}, integrate audio perception with large language models to support audio understanding, and achieve strong performance across a range of audio tasks, 
including automatic speech recognition~\citep{yang2024mala}, speech emotion understanding~\citep{wang2024blsp}, and audio captioning~\citep{chen2025slam}. 
However, their audio front ends are predominantly built on encoders such as Whisper~\citep{radford2023whisper}, WavLM~\citep{chen2022wavlm}, and EAT~\citep{chen2024eat}, which were developed for conventional human-audible audio, including speech, music, and environmental sounds\cmt. Consequently, the hearing capabilities of current ALLMs remain largely centered on the human-audible regime, while their ability to handle the broader and highly heterogeneous input regimes encountered in machine hearing has received limited study. Our work investigates this extension beyond human auditory perception, and further explores how existing ALLMs can be augmented for machine hearing while retaining their conventional audio capabilities.

\section{Machine Hearing Benchmark}
\label{sec:benchmark}

\begin{figure}[t]
  \centering
    \centering
    \vspace{-2.6em}
    \includegraphics[width=0.92\linewidth]{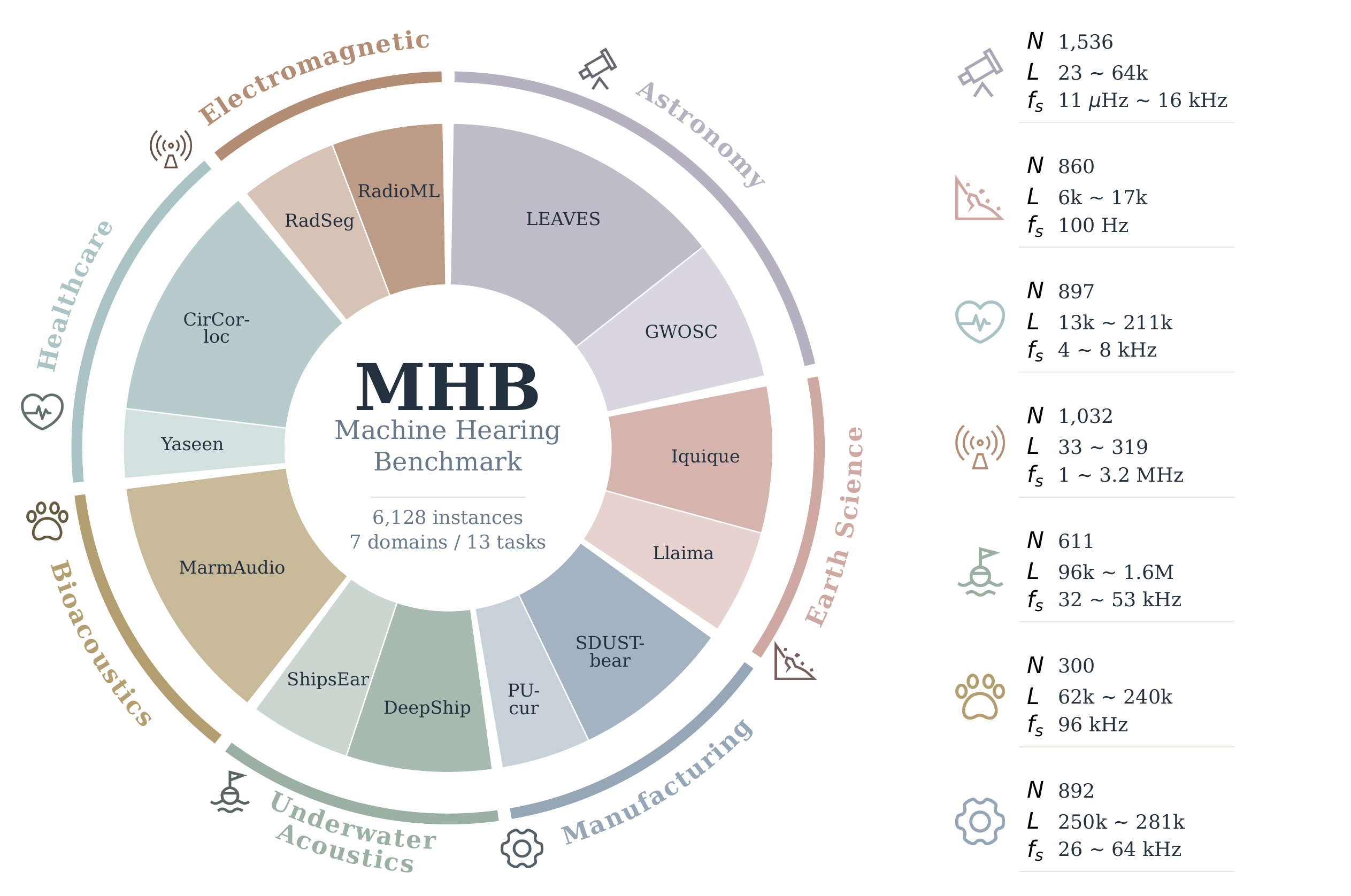}
    \vspace{-2mm}
    \caption{
    Overview of the Machine Hearing Benchmark (MHB), comprising 6,128 instances across 13 tasks and seven domains. For each domain, $N$, $L$, and $f_s$ denote the number of instances, sequence length, and sampling rate, respectively.
    }
  \hfill
  \label{fig:mhb-composition}
  \vspace{-1.8em}
\end{figure}

% llw
The Machine Hearing Benchmark (MHB) comprises 6,128 instances across 13 tasks from seven domains: Astronomy, Earth Science, Healthcare, Electromagnetic, Underwater Acoustics, Bioacoustics, and Manufacturing. As summarized in Fig.~\ref{fig:mhb-composition}, MHB spans sampling rates from approximately \(10^{-5}\) to \(10^{6}\) Hz, sequence lengths spanning approximately \(10^{1}\) to \(10^{6}\) samples, and single- and multi-channel signals.
Across domains, these signals differ by several orders of magnitude in sampling rate and sequence length and are acquired through markedly different sensing devices, highlighting the heterogeneity of MHB.

\subsection{Data Curation and Task Formulation}

\cmt
MHB is curated from 13 publicly available datasets and domain-specific data sources, supplemented by numerical simulations where appropriate. For each task, we retain the original input measurements and task semantics, with target labels derived from the corresponding annotations or metadata. We select tasks that can be naturally formulated as machine-hearing problems while preserving their original scientific or application objectives. 
Details about the data sources can be found in Appendix~\ref{apdx:dataset_detailed}.

To support unified evaluation with ALLMs and enable consistent evaluation across heterogeneous domains, each MHB instance consists of an input, a task-specific textual prompt, and a predefined label set. We preserve the original task definitions and label semantics while standardizing the prompt structure and response format across tasks, allowing different machine-hearing problems to be evaluated under a unified Audio LLM protocol. 
Representative prompt-response formulations are provided in Appendix~\ref{apdx:case}. 

% Finally, all data undergo systematic quality control to verify their consistency, correctness, and scientific validity.
% llw
% MHB is constructed from three primary sources: publicly available scientific datasets, domain-specific databases, and numerical simulations. Task construction is guided by the accompanying metadata of each dataset, which defines the input signal and its corresponding target label. To provide a unified evaluation interface across diverse sensing modalities, all tasks are formulated in a prompt-based format, ensuring compatibility with the training and evaluation paradigms of large language models (LLMs). Target labels are derived directly from the original metadata or expert annotations to preserve consistency with the underlying physical observations. Finally, all data undergo systematic quality control to verify their consistency, correctness, and scientific validity.

\subsection{Domains and Tasks}

% \begin{wraptable}{r}{0.45\textwidth}
% \vspace{-0.8em}
% \centering
% \caption{Statistics of MHB}
% \label{tab:mhb_stats}
% \begin{tabular}{lr}
% \toprule
% \textbf{Statistics} & \textbf{Number} \\
% \midrule
% \# Disciplines & 7 \\
% \# Domain tasks & 13 \\
% \midrule
% \# Instances & 6128 \\
% \quad \# Astronomy & 2104 \\
% \quad \# Earth Science & 1239 \\
% \quad \# Healthcare & 1521 \\
% \quad \# Electromagnetic & 1032 \\
% \quad \# Underwater Acoustics & 611 \\
% \quad \# Bioacoustics & 1229 \\
% \quad \# Manufacturing & 1228 \\
% \midrule
% Numerical Value Range & $10^{-7} \sim 10^{6}$ \\
% Frequency Range (Hz) & $10^{-5} \sim 10^{6}$ \\
% Dimension Range & $1 \sim 5$ \\
% Length Range & $10^{1} \sim 10^{6}$ \\
% \bottomrule
% \end{tabular}
% \vspace{-0.8em}
% \end{wraptable}

% llw
\cmt
The 13 tasks in MHB span seven domains and cover a broad range of waves, vibrations, oscillations, and acoustic phenomena. These include seismic waveforms, physiological recordings, bioacoustic and underwater acoustic signals, mechanical vibration and motor-current signals, radio-frequency and radar signals, gravitational-wave signals, and astronomical light curves.
% llw
% MHB covers seven scientific and engineering domains, encompassing a broad spectrum of sensor-recorded signal modalities beyond conventional audible sounds. The benchmark includes acoustic signals, infrasound, electromagnetic signals, gravitational waves, astronomical light curves, phonocardiograms (PCGs), vibration signals, current signals, radar signals, and radio-frequency signals, spanning diverse physical phenomena across different frequency ranges and sensing mechanisms.

% llw
In \textbf{Astronomy}, MHB includes gravitational-wave event detection and variable-star classification from stellar light curves. In \textbf{Earth Science}, MHB includes earthquake focal-depth classification from seismic waves and 
% volcanic activity classification from low-frequency infrasound. 
seismic event-type classification in volcanic environments. 
In \textbf{Healthcare}, MHB includes 
% fetal gestational age estimation and 
auscultation-location identification and 
cardiac disease classification from phonocardiogram recordings. In the \textbf{Electromagnetic} domain, MHB includes modulation scheme classification from radio-frequency waveforms and 
% pulse-pattern classification from radar signals. 
waveform-type classification from radar pulse segments. 
In \textbf{Underwater Acoustics}, 
% MHB includes ship-type classification from underwater recordings. 
MHB includes vessel and sound-source classification from underwater acoustic recordings. 
% of vessel sounds. 
In \textbf{Bioacoustics}, MHB includes 
% species-level 
marmoset vocalization-type classification. 
% , including primate and insect calls. 
% In \textbf{Manufacturing}, MHB includes bearing fault diagnosis and motor fault diagnosis from mechanical vibration, motor-current, and acoustic recordings. 
In \textbf{Manufacturing}, MHB includes bearing fault diagnosis from mechanical vibration and motor-current signals.
% , as well as machine-condition classification from recorded equipment operating sounds.
Detailed task-level statistics are provided in Appendix~\ref{apdx:task_detailed}.

\begin{figure}[t]
  \centering
    \centering
    \vspace{-2.2em}
    \includegraphics[width=\linewidth]{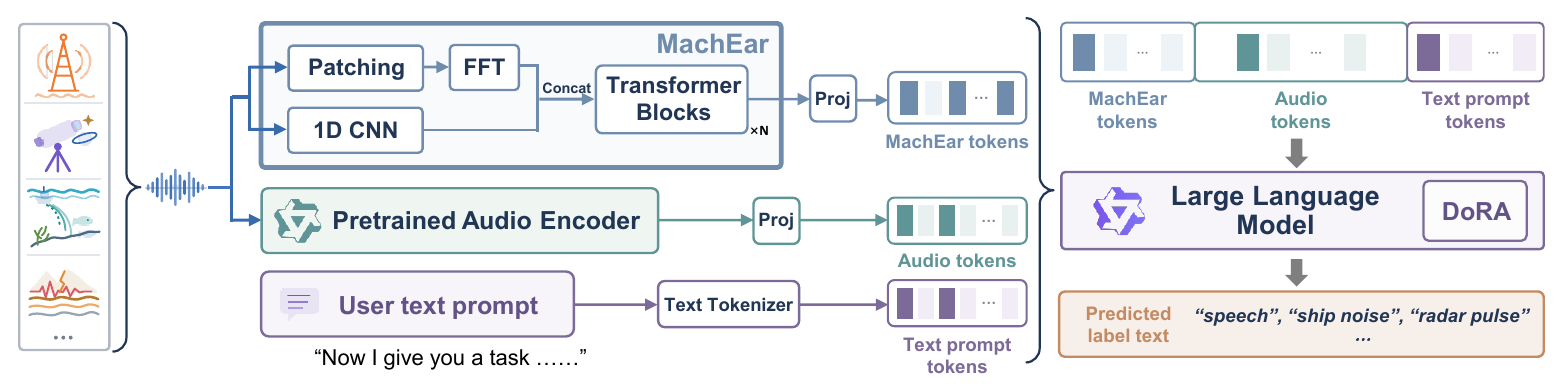}
    \vspace{-6mm}
    \caption{Overview of MachEar integrated with a pretrained audio pathway. Given an input signal, MachEar and the pretrained audio encoder process the same input in parallel to produce complementary representations. The resulting MachEar tokens, audio tokens, and text prompt tokens are concatenated and jointly fed into the LLM for prediction.}
  \hfill
  % MHB spans heterogeneous sensing domains, datasets, signal frequencies, and acquisition devices.}
  \label{fig:MachEar-overview}
  \vspace{-1.6em}
\end{figure}

\section{MachEar}
\label{sec:model}
% llw
The broad variation in machine-hearing inputs calls for an extension that can accommodate diverse input regimes while retaining the pretrained audio capabilities of existing Audio LLMs. We therefore introduce MachEar, a lightweight encoder extension that broadens the effective coverage of the pretrained audio encoder toward low- and high-frequency regimes and extremely short sequences. As illustrated in Fig.~\ref{fig:MachEar-overview}, MachEar processes the same input in parallel with the pretrained audio encoder, and the resulting MachEar tokens, audio tokens, and text prompt tokens are jointly fed into the LLM for prediction. 

MachEar combines spectral and temporal features to handle diverse machine-hearing inputs. Specifically, 
% FFT features computed over patched inputs are concatenated with features extracted by 1D CNN, followed by Transformer blocks and a projection layer that produces a compact sequence of MachEar tokens\cmt. 
% FFT and 1D CNN features are extracted from the input through two parallel pathways and concatenated, followed by Transformer blocks and a projection layer that encode the full input sequence into a fixed number \(Q\) of compact MachEar tokens. 
FFT and 1D CNN features are extracted from the input through two parallel pathways and then concatenated. The resulting features are processed by Transformer blocks and projection layers to produce a fixed number \(Q\) of compact MachEar tokens.
The same architecture is applied across all machine-hearing domains without domain-specific modifications.
More architectural and implementation details of MachEar are provided in Appendix~\ref{apdx:MachEar}.

\section{Experimental Setup}

\subsection{Baseline Models}

\begin{figure*}[t]
    \centering
    \vspace{-2.2em}
    \includegraphics[width=\textwidth]{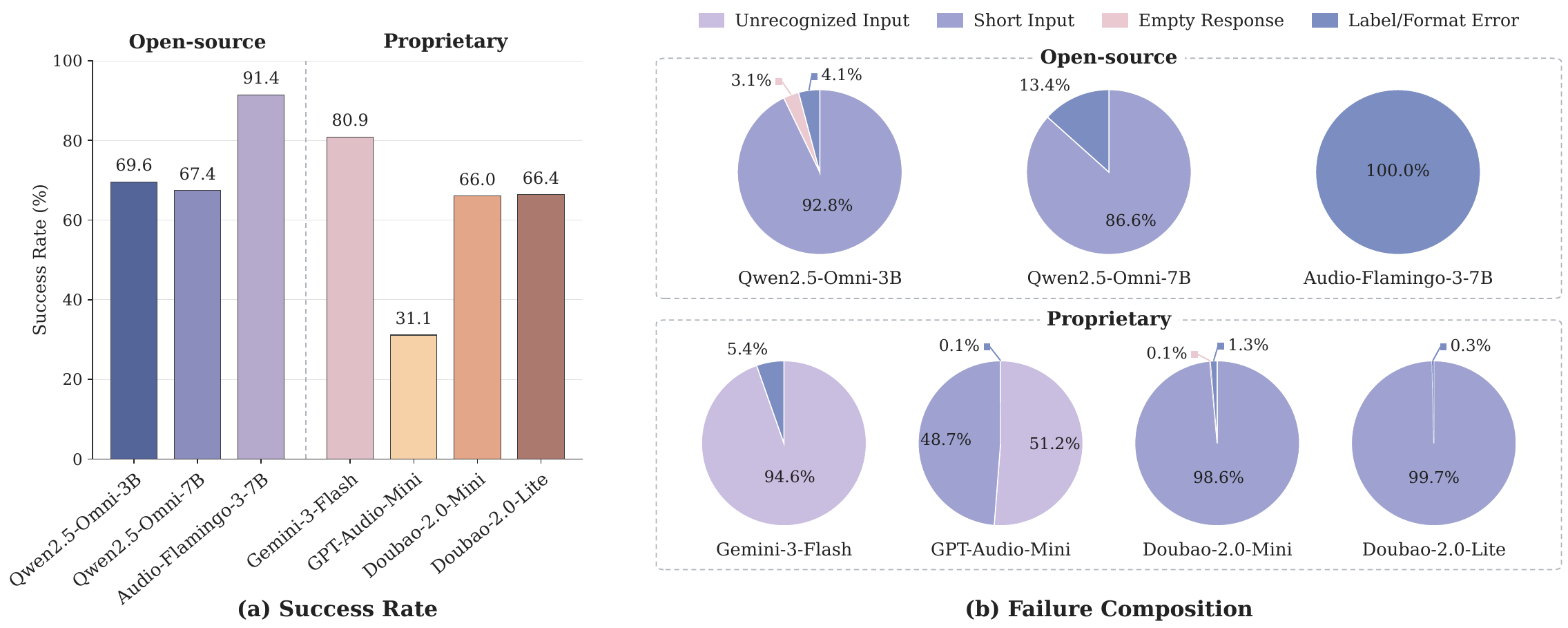}
    \caption{
    Response validity of Audio LLMs on MHB. (a) Overall success rate (SR), defined as the percentage of valid responses. (b) Composition of invalid responses across four failure categories.
    }
    \label{fig:sr&fc}
    \vspace{-1.0em}
\end{figure*}

We evaluate seven ALLMs, including three open-source and four proprietary models. To contextualize the performance of ALLMs, we also evaluate five foundation models pretrained on specialized signals from related domains. These models provide a complementary reference for examining how pretraining in related domains beyond speech transfers to machine-hearing tasks. 

\textbf{ALLMs.} We evaluate three open-source models, Qwen2.5-Omni-3B\footnote{\url{https://huggingface.co/Qwen/Qwen2.5-Omni-3B}}, Qwen2.5-Omni-7B\footnote{\url{https://huggingface.co/Qwen/Qwen2.5-Omni-7B}}~\citep{xu2025qwen25omni}, and Audio-Flamingo-3-7B\footnote{\url{https://huggingface.co/nvidia/audio-flamingo-3}}~\citep{ghosh2025audioflamingo3}, together with four proprietary models: Gemini-3-Flash\footnote{\url{https://ai.google.dev/gemini-api/docs/models/gemini-3-flash-preview}}, GPT-Audio-Mini\footnote{\url{https://developers.openai.com/api/docs/models/gpt-audio-mini}}, Doubao-2.0-Mini\footnote{\label{fn:doubao}\url{https://www.volcengine.com/product/doubao/}}, and Doubao-2.0-Lite\textsuperscript{\hyperref[fn:doubao]{\ref*{fn:doubao}}}. All Audio LLMs are evaluated zero-shot on MHB using a unified prompting protocol. \cmt

% llw
\textbf{Domain-specific Foundation Models.} We additionally evaluate five pretrained foundation models from related domains: Time-MoE\footnote{\url{https://huggingface.co/Maple728/TimeMoE-200M}}~\citep{shi2025timemoe} for general time series, SpectrumFM\footnote{\url{https://github.com/ChunyuLiu188/SpectrumFM}}~\citep{zhou2025spectrumfm} for radio spectrum, SeisLM\footnote{\url{https://github.com/liutianlin0121/seisLM}}~\citep{liu2024seislm} for seismic waveforms, FISHER\footnote{\url{https://huggingface.co/jiangab/FISHER-small-0723}}~\citep{fan2026fisher} for industrial signals, and SPEAR\footnote{\url{https://huggingface.co/marcoyang/spear-large-speech-audio}}~\citep{yang2025spear} for speech and audio. For each MHB task, each foundation model is kept frozen, and 5-nearest-neighbor classification is performed in its representation space using the corresponding training set as the reference set. 
Detailed model descriptions and experimental setup are provided in Appendix~\ref{apdx:model_access}.

% llw

% llw
\subsection{Evaluation Metrics}
We evaluate model performance using Macro-F1 and Success Rate (SR). Macro-F1 is used as the primary classification metric to account for class imbalance across tasks. 
% Success Rate measures the percentage of evaluated samples for which an ALLM produces a valid and parseable response within the predefined task-specific label set. 
Success Rate measures the percentage of evaluated instances for which an ALLM produces a valid and parseable response within the predefined task-specific label set. 
Representative invalid responses and the corresponding failure categories are provided in Appendix~\ref{apdx:failure_modes}.

\section{Results and Discussion}
\label{sec:results}
% We organized the results around the progression from existing baselines to the
% proposed model. We first benchmark Audio LLMs and domain-specific foundation models
% (FMs), then test whether supervised fine-tuning (SFT) alone closes the observed gap.
% We next compare our full model with both groups of baselines and finally isolate the
% effect of each architectural component. Unless otherwise stated, classification
% performance is measured by macro-F1 on the same MHB splits and label spaces.

\subsection{Benchmarking ALLMs on MHB}
\label{sec:results_baselines}

\begin{table*}[t]
\centering
\vspace{-2.0em}
\caption{
% Zero-shot performance of Audio LLMs on the 13 MHB tasks. All task-level results are reported as Macro-F1 (\%), with invalid responses treated as incorrect predictions. Random Guess is included as a reference baseline.
Zero-shot Macro-F1 (\%) of Audio LLMs on MHB, with invalid responses treated as incorrect predictions. Random Guess is included as a reference.
}
\label{tab:baseline}
\vspace{-0.4em}
\setlength{\tabcolsep}{3.5pt}
\renewcommand{\arraystretch}{1.06}

\resizebox{\textwidth}{!}{%
\begin{tabular}{@{}llccccccc@{}}

% =========================================================
% Upper part
% =========================================================
\toprule

\multirow{2}{*}{\textbf{Access}}
&
\multirow{2}{*}{\textbf{Model}}
&
\multicolumn{2}{c}{\textbf{Astronomy}}
&
\multicolumn{2}{c}{\textbf{Earth Science}}
&
\multicolumn{3}{c}{\textbf{Healthcare}}
\\

\cmidrule(lr){3-4}
\cmidrule(lr){5-6}
\cmidrule(lr){7-9}

& &
\textbf{GWOSC}
&
\textbf{LEAVES}
&
\textbf{Iquique}
&
\textbf{Llaima}
&
\textbf{Yaseen}
&
\multicolumn{2}{c}{\textbf{CirCor-loc}}
\\

\midrule

--
&
Random Guess
&
49.95
&
8.94
&
40.66
&
22.64
&
19.79
&
\multicolumn{2}{c}{24.92}
\\

\midrule

\multirow{3}{*}{\textbf{Open-source}}
&
Qwen2.5-Omni-3B
&
34.36
&
1.87
&
% 18.57
18.56
&
5.91
&
7.87
&
% \multicolumn{2}{c}{23.81}
\multicolumn{2}{c}{22.30}
\\

&
Qwen2.5-Omni-7B
&
38.11
&
0.00
&
% 25.46
24.78
&
3.75
&
% 12.60
12.06
&
% \multicolumn{2}{c}{22.06}
\multicolumn{2}{c}{18.65}
\\

&
Audio-Flamingo-3-7B
&
49.83
&
0.22
&
34.77
&
3.75
&
7.74
&
\multicolumn{2}{c}{17.83}
\\

\midrule

\multirow{4}{*}{\textbf{Proprietary}}
&
Gemini-3-Flash
&
35.62
&
0.00
&
31.88
&
% 23.85
23.16
&
16.05
&
\multicolumn{2}{c}{15.68}
\\

&
GPT-Audio-Mini
&
% 6.82
2.75
&
0.00
&
% 15.70
15.64
&
% 6.69
9.14
&
16.96
&
% \multicolumn{2}{c}{18.95}
\multicolumn{2}{c}{18.57}
\\

&
Doubao-2.0-Mini
&
32.73
&
0.00
&
31.02
&
14.89
&
7.74
&
\multicolumn{2}{c}{11.91}
\\

&
Doubao-2.0-Lite
&
33.10
&
0.00
&
% 31.51
31.36
&
16.47
&
7.74
&
\multicolumn{2}{c}{21.53}
\\

\midrule
\midrule

% =========================================================
% Lower part
% =========================================================

\multirow{3}{*}{\textbf{Access}}
&
\multirow{3}{*}{\textbf{Model}}
&
\multicolumn{2}{c}{\textbf{Electromagnetic}}
&
\multicolumn{2}{c}{\textbf{Underwater Acoustics}}
&
\multicolumn{1}{c}{\textbf{Bioacoustics}}
&
\multicolumn{2}{c}{\textbf{Manufacturing}}
\\

\cmidrule(lr){3-4}
\cmidrule(lr){5-6}
\cmidrule(lr){7-7}
\cmidrule(lr){8-9}

& &
\textbf{RadSeg}
&
\textbf{RadioML}
&
\textbf{ShipsEar}
&
\textbf{DeepShip}
&
\makecell{\textbf{Marm}\\\textbf{Audio}}
&
\makecell{\textbf{SDUST-}\\\textbf{bear}}
&
\makecell{\textbf{PU-}\\\textbf{cur}}
\\

\midrule

--
&
Random Guess
&
19.96
&
9.05
&
6.46
&
24.88
&
16.60
&
9.94
&
32.54
\\

\midrule

\multirow{3}{*}{\textbf{Open-source}}
&
Qwen2.5-Omni-3B
&
6.87
&
0.00
&
0.54
&
8.69
&
5.35
&
1.76
&
18.18
\\

&
Qwen2.5-Omni-7B
&
8.25
&
0.00
&
1.11
&
8.76
&
% 9.66
9.70
&
3.44
&
20.81
\\

&
Audio-Flamingo-3-7B
&
6.64
&
1.52
&
0.83
&
0.00
&
6.09
&
4.84
&
% 27.16
23.16
\\

\midrule

\multirow{4}{*}{\textbf{Proprietary}}
&
Gemini-3-Flash
&
10.60
&
% 6.59
6.11
&
8.92
&
16.26
&
% 14.52
13.77
&
11.46
&
27.51
\\

&
GPT-Audio-Mini
&
0.00
&
0.00
&
5.57
&
5.76
&
0.00
&
4.50
&
% 25.11
14.75
\\

&
Doubao-2.0-Mini
&
0.00
&
0.00
&
15.98
&
12.82
&
17.85
&
3.64
&
14.15
\\

&
Doubao-2.0-Lite
&
0.00
&
0.00
&
8.91
&
16.80
&
19.51
&
2.21
&
11.26
\\

\bottomrule

\end{tabular}%
}
\vspace{-0.5em}
\end{table*}

\textbf{Response Validity and Failure Modes.} We first examine whether current ALLMs can reliably process MHB inputs and produce valid responses. The overall success rate of each model and the breakdown of its typical failure modes are reported in Fig.~\ref{fig:sr&fc}. Although success rates vary across models, invalid responses remain non-negligible across the board, indicating that reliable input handling and task-compliant generation are themselves part of the challenge on MHB. 
% A notable fraction of invalid responses arises from short-input failures, which are concentrated in tasks with atypical signal regimes, including the ultra-low-sampling-rate LEAVES task and the MHz-range RadSeg and RadioML tasks. 
Short-input failures account for a substantial proportion of invalid responses for several models, particularly Qwen2.5-Omni-3B and Qwen2.5-Omni-7B. Task-level results in Appendix~\ref{apdx:failure_modes} further reveal frequent invalid responses on tasks with atypical signal regimes, including the ultra-low-sampling-rate LEAVES task and the MHz-range RadSeg and RadioML tasks. 
% Beyond this shared pattern, failure modes remain strongly model dependent: Gemini-3-Flash is dominated by unrecognized input 
% where 
% often treating the provided input as absent and requesting the time-series, I/Q samples, or acoustic features\cmt , while Audio Flamingo 3 primarily produces label or format violations that arise from responses outside the predefined label set, such as ``Tank'' instead of ``Tanker''\cmt. 
Failure modes also vary considerably across models: Gemini-3-Flash's invalid responses are dominated by unrecognized-input failures, in which the model often treats the provided signal as absent and requests the time-series data, I/Q samples, or acoustic features, while Audio Flamingo 3's invalid responses consist of label or format violations, such as ``Tank'' instead of ``Tanker''. 
% or a gestational-age range rather than a single class
GPT-Audio-Mini shows the lowest success rate, with its invalid responses almost evenly split between unrecognized-input and short-input failures. 
% A detailed task-level breakdown of success rates and r
Representative invalid responses are provided in Appendix~\ref{apdx:failure_modes}.

\textbf{Performance.} Table~\ref{tab:baseline} reports task-level Macro-F1 \cmt results on MHB. 
Macro-F1 is computed over all MHB instances, with invalid responses treated as incorrect predictions. 
Even when taking the best ALLM result for each task, performance remains below random guessing on 9 of the 13 tasks. Particularly low scores are observed on tasks that combine extreme sampling rates with short sequences, including the ultra-low-rate LEAVES task and the MHz-range RadSeg and RadioML tasks. Notably, these are also among the tasks with frequent input-handling failures in the preceding success-rate analysis. 
Even when the models produce valid responses, their predictions remain largely uninformative. At the same time, relative strengths vary across models and domains, with no single ALLM consistently dominating MHB. 
Overall, these results show that current ALLMs struggle to generalize beyond human auditory perception, with performance remaining sensitive to both signal characteristics and task domain.

\subsection{Benchmarking Domain Foundation Models on MHB}
\label{sec:domain-fm-results}
Beyond ALLMs, we further evaluate foundation models pretrained on related domains to provide a complementary view of MHB tasks. Table~\ref{tab:baseline_fms} summarizes their performance. First, results show that domain-specific pretraining can provide a strong advantage when the pretraining and target tasks are well aligned, as expected. For example, FISHER achieves the best performance on both Manufacturing tasks. However, this advantage is not universal: SeisLM leads on Iquique but not Llaima, and SpectrumFM does not rank first on either Electromagnetic task. Conversely, strong transfer is not restricted to nominally matched domains. SPEAR, an audio foundation model, remains competitive across many tasks outside conventional audio, while the general time-series model Time-MoE leads on LEAVES and RadSeg. Overall, these results suggest that domain specialization alone does not guarantee consistently strong transfer across heterogeneous MHB tasks.

\subsection{Can Supervised Adaptation Close the Machine-Hearing Gap?}
\label{sec:results_sft}

\begin{table*}[t]
\centering
\vspace{-2.2em}
% \caption{Macro-F1 (\%) of domain-specific foundation models on MHB using frozen representations and 5-NN classification. Best and second-best results are shown in bold and underlined, respectively.
% }
\caption{Macro-F1 (\%) of pretrained foundation models on MHB using frozen representations and 5-NN classification. Best and second-best results are shown in bold and underlined, respectively.
}
\label{tab:baseline_fms}
\vspace{-0.4em}
\setlength{\tabcolsep}{6.5pt}
\renewcommand{\arraystretch}{1.05}
\resizebox{0.97\textwidth}{!}{%
\begin{tabular}{@{}lccccccc@{}}

% =========================================================
% Upper domains
% =========================================================
\toprule

\multicolumn{1}{l}{\multirow{2}{*}{\textbf{Model}}} &
\multicolumn{2}{c}{\textbf{Astronomy}} &
\multicolumn{2}{c}{\textbf{Earth Science}} &
\multicolumn{3}{c}{\textbf{Healthcare}} \\
\cmidrule(lr){2-3}
\cmidrule(lr){4-5}
\cmidrule(lr){6-8}

&
\textbf{GWOSC} &
\textbf{LEAVES} &
\textbf{Iquique} &
\textbf{Llaima} &
\textbf{Yaseen} &
\multicolumn{2}{c}{\textbf{CirCor-loc}} \\

\midrule

Time-MoE
& 59.18
& \textbf{67.07}
& 46.82
& \underline{94.11}
& 93.09
& \multicolumn{2}{c}{27.87} \\

SpectrumFM
& 52.86
& \underline{61.70}
& 48.85
& 79.48
& 76.05
& \multicolumn{2}{c}{26.52} \\

SeisLM
& 48.75
& 58.42
& \textbf{60.42}
& 93.33
& 93.08
& \multicolumn{2}{c}{29.12} \\

FISHER
& \underline{68.75}
& 0.00
& 0.00
& 0.00
& \textbf{94.20}
& \multicolumn{2}{c}{\underline{38.57}} \\

SPEAR
& \textbf{78.91}
& 44.82
& \underline{52.28}
& \textbf{95.51}
& \underline{93.88}
& \multicolumn{2}{c}{\textbf{51.07}} \\
\midrule
\midrule
% =========================================================
% Lower domains
% =========================================================

\multicolumn{1}{l}{\multirow{2}{*}{\textbf{Model}}} &
\multicolumn{2}{c}{\textbf{Electromagnetic}} &
\multicolumn{2}{c}{\textbf{Underwater Acoustics}} &
\multicolumn{1}{c}{\textbf{Bioacoustics}} &
\multicolumn{2}{c}{\textbf{Manufacturing}} \\
\cmidrule(lr){2-3}
\cmidrule(lr){4-5}
\cmidrule(lr){6-6}
\cmidrule(lr){7-8}

&
\textbf{RadSeg} &
\textbf{RadioML} &
\textbf{ShipsEar} &
\textbf{DeepShip} &
\makecell{\textbf{Marm}\\\textbf{Audio}} &
\makecell{\textbf{SDUST-}\\\textbf{bear}} &
\makecell{\textbf{PU-}\\\textbf{cur}} \\

\midrule

Time-MoE
& \textbf{88.12}
& 2.82
& 7.62
& 20.96
& 15.61
& 22.80
& 36.08 \\

SpectrumFM
& 82.92
& \underline{44.19}
& 13.68
& 32.10
& 22.45
& 32.26
& 21.51 \\

SeisLM
& \underline{87.22}
& \textbf{45.33}
& 20.08
& \underline{49.98}
& 29.11
& 55.09
& 33.42 \\

FISHER
& 0.00
& 0.00
& \underline{23.17}
& 48.69
& \textbf{92.57}
& \textbf{73.93}
& \textbf{53.69} \\

SPEAR
& 67.42
& 31.02
& \textbf{25.96}
& \textbf{58.69}
& \underline{77.55}
& \underline{60.57}
& \underline{53.34} \\

\bottomrule

\end{tabular}%
}
% \vspace{-1.2em}
\end{table*}

This section investigates whether the poor performance of current ALLMs on MHB can be explained simply by a mismatch between their training data and machine-hearing domains.
We therefore examine whether direct supervised adaptation on machine-hearing data is sufficient to close the gap. 
% We take Qwen2.5-Omni-7B as the backbone, finetune the LLM with DoRA, and compare two configurations: \texttt{Whisper\_frz}, where the pretrained Whisper encoder is frozen, and \texttt{Whisper}, where it is jointly finetuned on MHB training data. 
We take Qwen2.5-Omni-7B, whose audio encoder is initialized from Whisper-large-v3, as the backbone, finetune the LLM with DoRA on machine-hearing adaptation data, and compare two configurations: \texttt{Whisper\_frz}, where the audio encoder is frozen, and \texttt{Whisper}, where it is jointly finetuned with the LLM. 
As shown in Fig.~\ref{fig:mh_ft}, supervised adaptation enables the model to make meaningful predictions on tasks where zero-shot ALLMs largely fail. Further finetuning the Whisper encoder improves performance on most tasks, showing that the pretrained audio front end itself needs to adapt to machine-hearing inputs. However, this adaptation does not resolve all limitations. Tasks such as LEAVES, RadSeg, and RadioML remain particularly challenging despite finetuning, and are characterized by sampling rates and sequence lengths far outside conventional audio regimes. Their persistent difficulty after task-specific adaptation suggests that the bottleneck is not merely data-domain mismatch, but architectural constraints of conventional audio front ends in handling the diverse input regimes of machine hearing. This motivates an architectural extension of the audio front end, which we pursue with MachEar in the next section.

\subsection{Augmenting ALLMs with MachEar}
To address the architectural limitations identified above, we introduce {MachEar}, a lightweight encoder extension for augmenting ALLMs with stronger machine-hearing capability. Our analysis of domain foundation models in Sec.~\ref{sec:domain-fm-results} provides useful architectural clues. Models that remain effective on challenging MHB tasks often adopt explicit frequency-domain processing or convolutional modeling of temporal inputs, suggesting inductive biases that are better suited to input regimes beyond conventional audio. Guided by these clues, MachEar combines FFT-based spectral modeling with a 1D CNN to complement the pretrained Whisper encoder. Rather than replacing the pretrained audio encoder, MachEar operates alongside it, broadening the effective coverage of the audio front end toward low- and high-frequency regimes and extremely short sequences.

\begin{figure*}[t]
    \centering
    \vspace{-0.6em}
    \includegraphics[width=\textwidth]{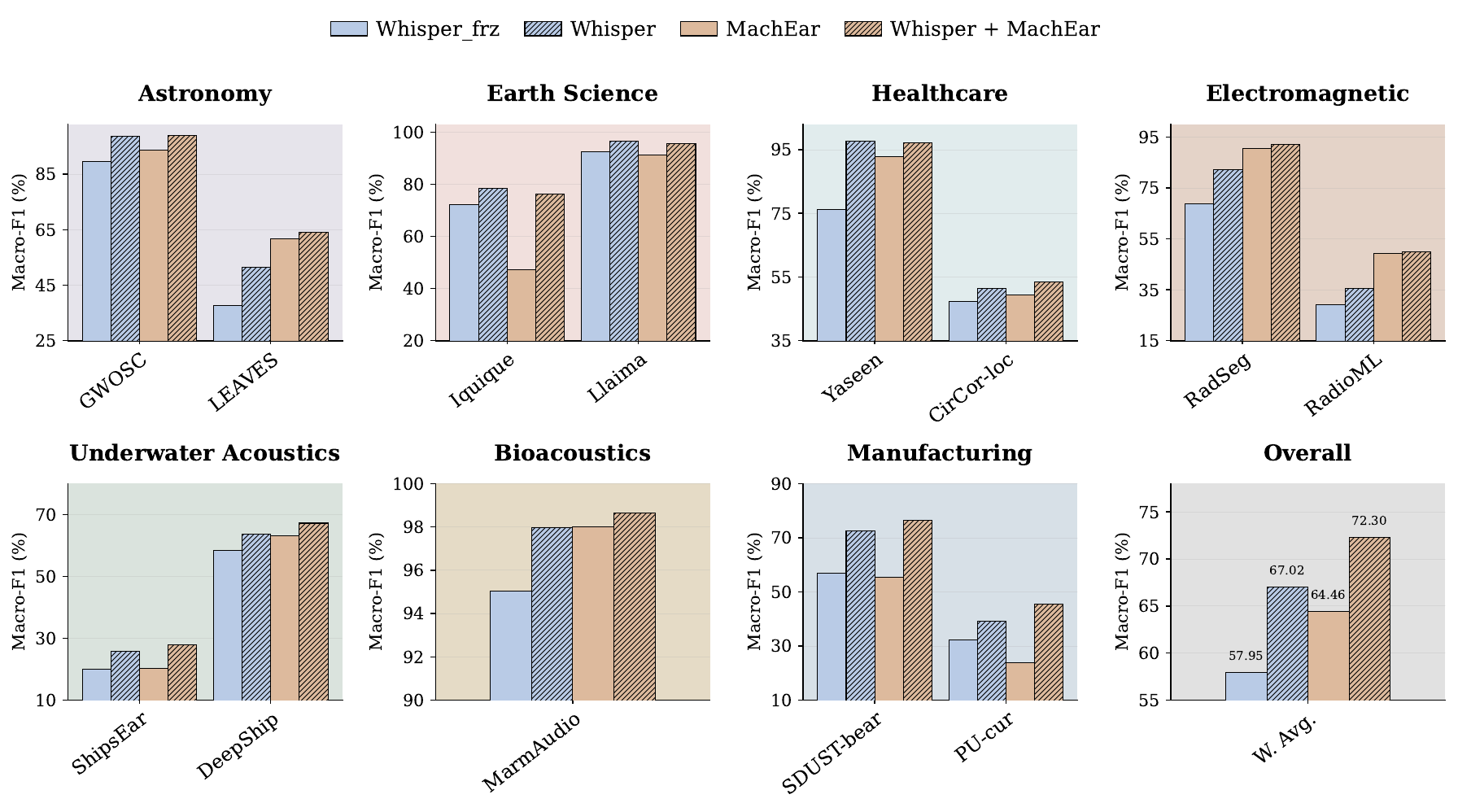}
    \vspace{-1.8em}
    \caption{
    % Task-level Macro-F1 (\%) on the 13 MHB tasks. The final column reports the test-set-size-weighted average across tasks. \cmt
    Task-level Macro-F1 (\%) of four adaptation configurations on MHB. All use DoRA for LLM adaptation: \texttt{Whisper\_frz} freezes Whisper; \texttt{Whisper} finetunes Whisper; \texttt{MachEar} replaces Whisper; and \texttt{Whisper + MachEar} jointly finetunes both encoders. \textit{W. Avg.} denotes the test-set-size-weighted average.
    }
    \label{fig:mh_ft}
    \vspace{-0.7em}
\end{figure*}

We evaluate two MachEar-based configurations using the same Qwen2.5-Omni-7B backbone: \texttt{MachEar}, which directly replaces the pretrained Whisper encoder with MachEar; and \texttt{Whisper + MachEar}, which adds MachEar to the Qwen backbone and jointly finetunes both encoders. 
As shown in Fig.~\ref{fig:mh_ft}, \texttt{MachEar} alone outperforms the finetuned \texttt{Whisper} configuration on a substantial subset of MHB tasks, especially LEAVES and RadSeg, 
% that have shorter length\cmt.
which involve shorter input sequences. 
This indicates that its frequency-domain and convolutional design captures signal structure that is not fully modeled by audio front ends developed for conventional audio tasks.
Accordingly, augmenting Qwen2.5-Omni-7B with MachEar yields more consistent gains across tasks, raising the overall weighted Macro-F1 from 67.02\% to 72.30\%\cmt. Together, these results indicate that MachEar provides complementary machine-hearing capacity beyond direct adaptation of the pretrained audio encoder.

% We first compare Qwen and Qwen-WFT to assess the effect of adapting the pretrained audio encoder during supervised finetuning on machine-hearing data. Compared with Qwen, Qwen-WFT improves performance on most of the 13 in-domain tasks, indicating the benefit of adapting the pretrained audio encoder for machine-hearing tasks. Comparing Qwen-WFT with MachEar, neither consistently dominates across tasks; instead, the two exhibit complementary task-level strengths. Notably, MachEar shows its clearest advantages over Qwen-WFT on short-sequence tasks such as LEAVES, RadSeg, and RadioML; this pattern is examined further in Sec.~\ref{sec:results_ablation}. MachEar+WFT achieves the best or tied-best performance on most tasks and remains competitive on the others, with ShipsEar being the clearest exception where Qwen-WFT performs substantially better. Accordingly, MachEar+WFT achieves the highest weighted Macro-F1 of 74.63\%, improving upon Qwen-WFT at 65.83\%. These results suggest that MachEar and the pretrained audio encoder provide complementary representations, and that combining them yields the most consistent gains across machine-hearing tasks.

\subsection{Ablation Study and Analysis of MachEar}
\label{sec:results_ablation}

This section provides a more detailed analysis of where MachEar is most effective and how its main components contribute to its performance. We group the 13 MHB tasks according to sequence length and sampling rate, resulting in four input regimes: 
\begin{itemize}
    % \item G1 -- short high-rate inputs (signal length \(<1\mathrm{k}\), sampling rate \(>96\mathrm{kHz}\));
    % \item  G2 -- short-to-medium low-rate inputs (signal length \(1\mathrm{k}\)–\(10\mathrm{k}\), sampling rate \(<2\mathrm{kHz}\)); 
    \item G1 -- short high-rate inputs (signal length \(<10\mathrm{k}\), sampling rate \(>96\,\mathrm{kHz}\));
    \item  G2 -- short low-rate inputs (signal length \(<10\mathrm{k}\), sampling rate \(<2\,\mathrm{kHz}\)); 
    \item G3 -- long low-rate inputs (signal length \(>10\mathrm{k}\), sampling rate \(<2\,\mathrm{kHz}\));
    \item G4 -- long inputs within conventional audio sampling rates (signal length \(>10\mathrm{k}\), sampling rate \(2\)–\(96\,\mathrm{kHz}\)).
\end{itemize}
Based on this grouping, we compare Qwen2.5-Omni-7B with and without MachEar, and further analyze the contributions of MachEar's FFT and 1D CNN components.

\begin{figure}[t]
  \centering
    \centering
    \vspace{-2.2em}
    \includegraphics[width=0.95\textwidth]{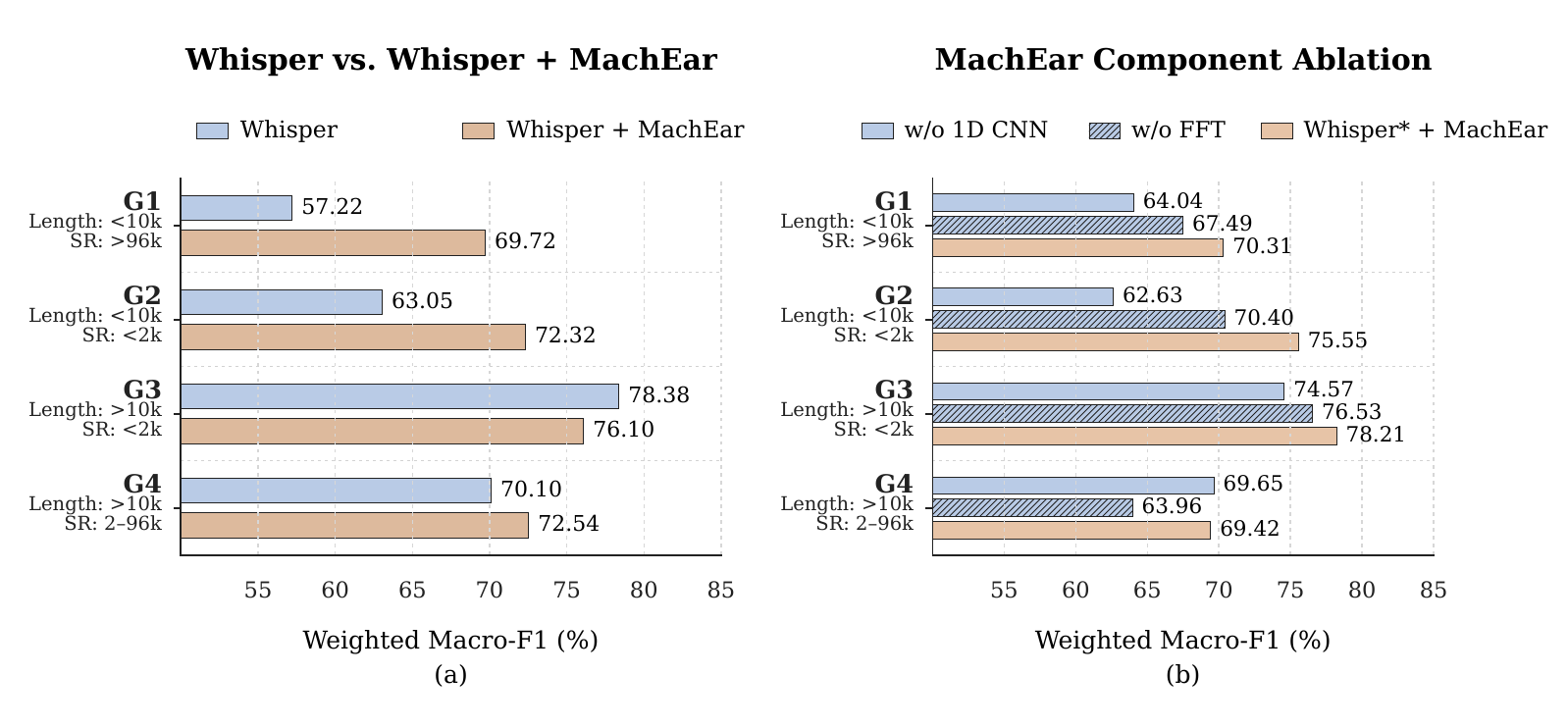}
    \vspace{-1.3em}
    \caption{
    % Analysis of MachEar across different signal characteristics. \cmt 
    % (a) The 13 MHB tasks are grouped into four regimes according to sequence length and sampling rate. 
    % (a) Comparison between Whisper and Whisper + MachEar across the four regimes. (b) Ablation of the FFT and 1D CNN components in MWhisper + MachEar. Results in (a) and (b) report weighted Macro-F1 (\%) within each group.
    MachEar analysis across four input regimes. (a) \texttt{Whisper} vs.\ \texttt{Whisper + MachEar}. (b) Component ablations, where * denotes a frozen pretrained Whisper encoder. Results are test-set-size-weighted Macro-F1 (\%) within each group.
    }
  \hfill
  % MHB spans heterogeneous sensing domains, datasets, signal frequencies, and acquisition devices.}
  \label{fig:MachEar-analysis}
  \vspace{-1.7em}
\end{figure}

\textbf{Effectiveness across Input Regimes.} We first examine how the benefit of augmenting Qwen2.5-Omni-7B with MachEar varies across the four input regimes by comparing \texttt{Whisper} with \texttt{Whisper + MachEar}. As shown in Fig.~\ref{fig:MachEar-analysis}(a)\cmt, adding MachEar provides the largest gains in G1 and G2, whereas its advantage becomes smaller for the longer-sequence G3 and G4 regimes. This pattern suggests that MachEar's complementary benefit is most pronounced for shorter sequences with sampling rates outside the conventional audio range, consistent with its design goal of compensating for the limitations of the original audio front end and extending the effective input range of ALLMs.

\textbf{Component Ablation.} We next examine the contributions of the two main components of MachEar across the four signal regimes. 
% To isolate the contribution of MachEar itself, we keep the pretrained Whisper encoder frozen across all ablation variants. 
To isolate the contribution of MachEar itself, we keep the pretrained Whisper encoder frozen across all ablation variants, with the full configuration denoted as \texttt{Whisper* + MachEar}. 
% As shown in Fig.~\ref{fig:MachEar-analysis}(b)\cmt, the 1D CNN contributes most clearly in the shorter-sequence regimes, with its removal reducing weighted Macro-F1 from 70.03\% to 66.18\% in G1 and from 78.04\% to 63.76\% in G2. In contrast, removing the FFT produces broader degradation across G2--G4, with drops of 7.38, 7.37, and 5.44 points, while having little effect on G1. These results reveal complementary roles of the two components of MachEar: the 1D CNN is particularly important for handling shorter sequences, whereas the FFT contributes more consistently across signals spanning a broader range of sampling rates.
As shown in Fig.~\ref{fig:MachEar-analysis}(b)\cmt, the 1D CNN pathway contributes most clearly in the shorter-sequence regimes, with its removal reducing weighted Macro-F1 from 70.31\% to 64.04\% in G1 and from 75.55\% to 62.63\% in G2. In contrast, removing the FFT pathway produces a broader degradation across all four regimes, with drops of 2.82, 5.15, 1.68, and 5.46  percentage points in G1--G4. These results reveal complementary roles of the two components of MachEar: the 1D CNN is particularly important for handling shorter sequences, whereas the FFT contributes more consistently across signals spanning a broader range of sampling rates.

% \subsection{Analysis of MachEar by Signal Characteristics}

% \begin{table*}[t]
% \centering
% \caption{Ablation study of the FFT and 1D CNN components in MachEar, with results aggregated by the maximum signal length (in samples) and sampling rate of each task. Values are macro-F1 (\%).}
% \label{tab:ablation}

% \resizebox{0.7\textwidth}{!}{%
% \begin{tabular}{lccc@{\hspace{0.5cm}}ccc}
% \toprule

% \multirow{2}{*}{\textbf{Variant}} &
% \multicolumn{3}{c}{\textbf{Signal Length (samples)}} &
% \multicolumn{3}{c}{\textbf{Sampling Rate (Hz)}} \\
% \cmidrule(lr){2-4} \cmidrule(lr){5-7} 
% &
% \textbf{$<$1k} &
% \textbf{1--10k} &
% \textbf{$>$10k} &
% \textbf{$<$2k} &
% \textbf{2--96k} &
% \textbf{$>$96k} \\
% \midrule

% Qwen2.5-Omni SFT
% & 55.46 & 56.01 & \textbf{72.36}
% & 62.99 & \underline{70.83} & 55.46 \\

% Full Model
% & \underline{70.03} & \textbf{78.04} & \underline{71.32}
% & \textbf{77.80} & 70.41 & \underline{70.03} \\

% w/o 1D CNN Pathway
% & 66.18 & 63.76 & 71.25
% & 66.27 & \textbf{70.94} & 66.18 \\

% w/o FFT Pathway
% & \textbf{70.97} & \underline{70.66} & 65.62
% & \underline{70.42} & 64.97 & \textbf{70.97} \\

% \bottomrule
% \end{tabular}%
% }
% \end{table*}

% llw
% To examine how the effectiveness of MachEar varies with signal characteristics, we group the 13 in-domain MHB tasks by sequence length and sampling rate, as shown in Fig.~\ref{fig:MachEar-analysis}(a). This yields four signal regimes: short high-rate signals (G1), short-to-medium low-rate signals (G2), long low-rate signals (G3), and long signals within the conventional audio sampling-rate range (G4). As defined in Table~\ref{tab:setup}, MachEar+W combines MachEar with a frozen pretrained audio encoder, while Qwen-WFT finetunes the pretrained audio encoder without MachEar.

% As shown in Fig.~\ref{fig:MachEar-analysis}(b), compared with Qwen-WFT, MachEar+W provides substantial gains in G1 and G2, while providing no improvement in the longer-sequence G3 and G4 regimes, suggesting that its complementary benefit is primarily observed for shorter sequences. Fig.~\ref{fig:MachEar-analysis}(c) further disentangles the contributions of the FFT and 1D CNN components across these four signal regimes. The 1D CNN contributes most clearly in the shorter-sequence regimes, with its removal reducing weighted Macro-F1 from 70.03\% to 66.18\% in G1 and from 78.04\% to 63.76\% in G2. The FFT pathway provides broader benefits across G2--G4, with its removal reducing weighted Macro-F1 by 7.38, 7.37, and 5.44 points, respectively, while having little effect in G1. Together, these results suggest complementary roles within MachEar: the 1D CNN is most beneficial for short sequences, whereas the FFT contributes more consistently across signals spanning a broader range of sampling rates.

\subsection{Preservation of Conventional Audio Capabilities}
\label{sec:results_audio}
\begin{wrapfigure}{r}{0.58\textwidth}
    \vspace{-1.2em}
    \centering
    \includegraphics[
        width=\linewidth
    ]{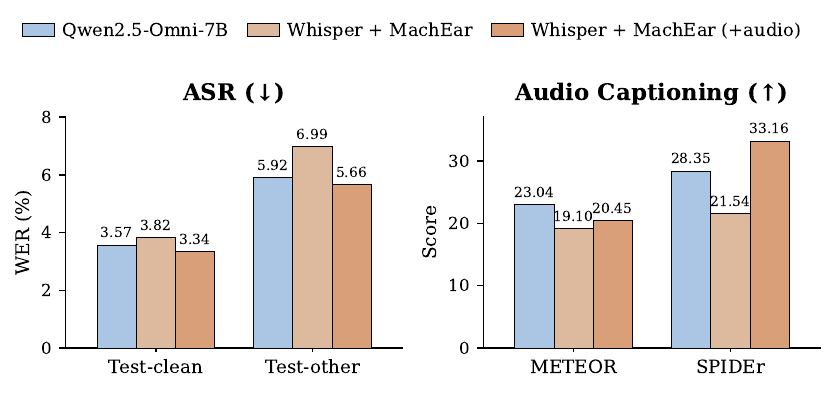}
    \vspace{-1.8em}
    \caption{
    Performance on conventional audio tasks under different adaptation configurations. Lower WER is better, while higher METEOR and SPIDEr are better.
    }
    \label{fig:ac_ft}
    \vspace{-1.0em}
\end{wrapfigure}

We next evaluate whether extending the model toward machine hearing affects its performance on conventional audio tasks. 
% We compare the machine-hearing-adapted \texttt{Whisper + MachEar} with the original Qwen2.5-Omni-7B on automatic speech recognition (ASR) and audio captioning tasks.
We compare the machine-hearing-adapted \texttt{Whisper + MachEar} with the original Qwen2.5-Omni-7B on automatic speech recognition (ASR) using the LibriSpeech test-clean and test-other sets, and on audio captioning using AudioCaps. 
As shown in Fig.~\ref{fig:ac_ft}, while degradation is expected after adapting the model to machine hearing, the observed drop on conventional audio tasks is surprisingly modest. The gap can be further narrowed by incorporating ASR and audio-captioning data during machine-hearing adaptation, 
as shown by \texttt{Whisper + MachEar (+audio)} \cmt in Fig.~\ref{fig:ac_ft}. 
A relatively small amount of conventional audio-task data, accounting for less than one third of the overall adaptation data, 
% is sufficient to largely recover the degradation and achieves performance comparable to, or even better than, the original Qwen2.5-Omni-7B across the evaluated metrics. 
is sufficient to largely restore conventional audio performance and achieve results comparable to, or even better than, those of the original Qwen2.5-Omni-7B across the evaluated metrics. 
Overall, these results suggest that MachEar can extend ALLMs toward machine hearing without substantially disrupting their original audio capabilities.

\section{Conclusion}
In this work, we investigate whether Audio LLMs can extend beyond human auditory perception toward a broader form of machine hearing. We introduce the Machine Hearing Benchmark (MHB), a diverse benchmark spanning 7 domains, 13 tasks, and highly heterogeneous input regimes, to systematically evaluate this capability. Benchmarking a broad set of current open-source and proprietary ALLMs reveals that current ALLMs struggle to generalize beyond human hearing. More importantly, supervised adaptation with task-specific machine-hearing data improves performance but does not fully close the gap, showing that the limitation cannot be explained by data-domain mismatch alone. Instead, the persistent failures point to architectural constraints in conventional audio front ends when handling the diverse input regimes encountered in broader machine hearing. Motivated by this finding, we introduce MachEar, a lightweight encoder extension that can be readily attached to existing ALLMs. Results show that MachEar complements the pretrained audio encoder, particularly for extremely short sequences and inputs at low- and high-sampling-rate extremes. Importantly, these gains can be achieved while largely preserving conventional audio performance. Overall, our findings expose the current limits of ALLMs while demonstrating their emerging potential to extend toward machine hearing. More broadly, they point toward multimodal LLMs whose perceptual scope is no longer bounded by human senses, enabling them to perceive and understand aspects of the physical world inaccessible to humans.

\section*{AI use statement}
LLMs were employed solely to polish the writing and correct grammatical errors.
\clearpage
% In this work, we study audio understanding from a machine-centric perspective, extending the scope from human-audible signals to broader sensor-recorded signals. We introduce the Machine Hearing Benchmark (MHB), which unifies 13 tasks across seven sensing domains. Evaluations reveal limited transfer of existing Audio LLMs beyond conventional audio, while scientific foundation models show stronger supervised performance but uneven cross-domain transfer.

% To bridge this gap, we propose MachEar, a lightweight encoder that complements the pretrained Whisper pathway with machine-hearing representations. Combined with a three-stage fine-tuning strategy, MachEar improves the weighted Macro-F1 of Qwen2.5-Omni from 63.20\% to 75.17\%, while largely preserving performance on the evaluated ASR and audio captioning tasks. Overall, MHB and MachEar provide a benchmark and modeling framework for extending Audio LLMs from human-centered auditory perception toward broader machine-hearing capabilities.

\cmt
\bibliography{iclr2026_conference}
\bibliographystyle{iclr2026_conference}

\newpage
\appendix

% \section{Extended Related work}

% \subsection{Audio Foundation Model}
% \textbf{Audio Encoder.}
% \textbf{Audio LLMs.}
% \textbf{Omni Models.}

% \subsection{Audio Benchmark and Scientific Benchmark}
% \textbf{Audio Benchmarks.}
% \textbf{Scientific Benchmarks.}

\section{More Details of MHB}
\subsection{Task Description}
\label{apdx:task_detailed}
Table~\ref{tab:mhb_task_stats} presents task-level statistics, including sequence-length and sampling-rate ranges and the corresponding data sources. We give a detailed description of each task below.

\begin{table}[t]
\caption{
% Task-level statistics of MHB.
Task-level statistics of MHB, including the number of instances, input sequence length, sampling rate, number of channels, and data source for each task.
}
\vspace{-0.4em}
\centering
\resizebox{\linewidth}{!}{
\setlength{\tabcolsep}{1pt}
\begin{tabular}{ccccccc}
\toprule
 \multirow{2}{*}{\textbf{Domain}}&\multirow{2}{*}{\textbf{Name}} &\multirow{2}{*}{\textbf{\# Instances}} &\textbf{Sample} & \textbf{Sampling}  &\textbf{Channel} &\multirow{2}{*}{\textbf{Data Source}}\\
  & & &\textbf{Length Range} &\textbf{Rate (Hz)} &\textbf{Range} & \\
\midrule
 \multirow{2}{*}{Astronomy} 
 &GWOSC&512 &6.4e4 &1.6e4 &1  &SciTS~\citep{wu2026scits}\\
 &LEAVES&1024 &23$\sim$1.3e3 &1.1e-5 &1 &LEAVES~\citep{fei2024leaves}\\
 % &StarEmbed&568 &71$\sim$4.5e3 &1.1e-5 &2 &StarEmbed~\citep{li2025starembed}\\
 \midrule
 \multirow{2}{*}{Earth Science}
 &Iquique&502 &1.4e4$\sim$1.7e4 &1.0e2 &3 &Iquique~\citep{woollam2019convolutional}\\
 &Llaima&358 &6.0e3 &1.0e2 &1 &Llaima~\citep{llaimadataset2020}\\
 % &Yasur&379 &2.5e2 &50 &5 &Yasur~\citep{toney2022yasur}\\
 \midrule
 \multirow{2}{*}{Healthcare}
 &Yaseen&200 &1.3e4$\sim$2.7e4 &8.0e3 &1 &Yaseen~\citep{yaseen2018classification}\\
 &CirCor-loc&697 &2.1e4$\sim$2.1e5 &4.0e3 &1 &CirCor~\citep{oliveira2021circor}\\
 % &PhysioNet&624 &3.3e3$\sim$3.4e3 &3.3e2 &1 &PhysioNet FPCG~\citep{cesarelli2012simulation}\\
 \midrule
 \multirow{2}{*}{Electromagnetic}
 &RadSeg&482 &33$\sim$3.2e2 &3.2e6 &2 &RadSeg~\citep{huang2024radseg}\\
 &RadioML&550 &1.3e2 &1.0e6 &2 &RadioML 2016.10A~\citep{o2016radio}\\
 \midrule
 \multirow{2}{*}
 {\headcell{Underwater\\Acoustics}}
 &ShipsEar&84 &1.6e6 &5.3e4 &1 &ShipsEar~\citep{santos2016shipsear}\\
 &DeepShip&527 &9.6e4 &3.2e4 &1 &DeepShip~\citep{irfan2021deepship}\\ 
 \midrule
 \multirow{1}{*}{Bioacoustics}
 &MarmAudio&300 &6.2e4$\sim$2.4e5 &9.6e4 &1 &MarmAudio~\citep{lamothe_2025_15017207}\\
 % &BEANS-HumBugDB&929 &5.0e2$\sim$4.0e6 &8.0e3$\sim$4.4e4 &1 &BEANS~\citep{hagiwara2023beans,kiskin2021humbugdb}\\
 \midrule
 \multirow{2}{*}{Manufacturing}
 &SDUST-bear&572 &2.6e5 &2.6e4 &1 &SDUST~\citep{han2022novel,fan2026fisher}\\
 &PU-cur&320 &2.5e5$\sim$2.8e5 &6.4e4 &1 &PU~\citep{lessmeier2016bearing}\\
 % &IIEE&336 &1.3e5 &4.4e4 &1 &IIEE~\citep{grollmisch2019sounding}\\
 \midrule
 \multirow{1}{*}{Overall}
 &-&6,128 &23$\sim$1.6e6 &1.1e-5$\sim$3.2e6 &1$\sim$3 &-\\
\bottomrule
\end{tabular}
}
\label{tab:mhb_task_stats}
\end{table}

All tasks in MHB are formulated as classification problems: each model receives a sensor-recorded signal together with a textual instruction and generates the corresponding class label in text form.

\textbf{Astronomy.} Astronomy studies celestial objects and physical phenomena in the universe. Two tasks are involved:
\begin{itemize}
    \item \textbf{GWOSC:}
    This task detects the presence of gravitational-wave events in time-series signals. It is formulated as a binary classification task, with ``normal'' indicating the absence of an event and ``anomaly'' indicating its presence. 
    \item \textbf{LEAVES:}
    This task identifies the variability category of a celestial object based on its brightness variations over time. It distinguishes seven categories: Cepheid, Delta Scuti, Eclipsing Binary, Long-Period Variable, Non-variable, Rotational Variable, and RR Lyrae. 
    % \item \textbf{StarEmbed:}
    % This task identifies stellar variability types through light-curve classification. It distinguishes seven categories: EW (W Ursae Majoris-type), EA (Algol-type), RRab (RR Lyrae, fundamental mode), RRc (RR Lyrae, first-overtone mode), RRd (RR Lyrae, double-mode), RS CVn (RS Canum Venaticorum) and  LPV (Long Period Variable).
\end{itemize}

\textbf{Earth Science.} Earth science studies the structure of Earth and its physical
processes. Two tasks are involved:
\begin{itemize}
    \item \textbf{Iquique:}
    This task classifies the focal depth of an earthquake source based on seismic recordings. It distinguishes three categories: shallow (0--70~km), intermediate (70--300~km), and deep (300--700~km). 
    % \item \textbf{Yasur:}
    % This task identifies the subcrater of origin of a volcanic explosion based on its infrasound signal. It is formulated as a binary classification, with \texttt{S} indicating the south subcrater and \texttt{N} indicating the north subcrater.
    \item \textbf{Llaima:}
    This task identifies seismic event types in a volcanic environment. It distinguishes four categories: Volcano-tectonic, Tremor, Tectonic, and Long-period. 
\end{itemize}

\textbf{Healthcare.} Healthcare focuses on assessing and maintaining human health. Two tasks are involved:
\begin{itemize}
    \item \textbf{Yaseen:}
    This task classifies heart sound patterns based on phonocardiogram (PCG) recordings. It distinguishes five categories: Aortic Stenosis, Mitral Regurgitation, Mitral Stenosis, Mitral Valve Prolapse, and Normal. 
    \item \textbf{CirCor-loc:}
    This task identifies the auscultation location from a PCG recording. It distinguishes four locations: aortic valve, tricuspid valve, pulmonary valve, and mitral valve. 
    % \item \textbf{PhysioNet FPCG:}
    % This task estimates gestational age from fetal phonocardiogram (fPCG) recordings. The target gestational ages range from 31 to 40 weeks.
\end{itemize}

\textbf{Electromagnetic.} Electromagnetic signal analysis studies electromagnetic waveforms and their applications in sensing and communication. Two tasks are involved:
\begin{itemize}
    \item \textbf{RadSeg:}
    This task identifies the waveform type of a pulsed radar signal segment. It distinguishes five categories: Barker codes, Polyphase Barker codes, Frank codes, Linear Frequency-Modulated (LFM) pulses, and Coherent Unmodulated Pulse Trains. 
    \item \textbf{RadioML:}
    % This task identifies the modulation type of radio signals from their I/Q waveforms. It distinguishes 11 analog and digital modulation schemes, including PSK, QAM, FSK, PAM, AM, and FM variants.
    This task identifies the modulation type of radio signals from their I/Q waveforms. It distinguishes 11 analog and digital modulation schemes: 8PSK, AM-DSB, AM-SSB, BPSK, CPFSK, GFSK, PAM4, QAM16, QAM64, QPSK, and WBFM. 
\end{itemize}

\textbf{Underwater Acoustics.} Underwater acoustics studies sound and acoustic phenomena in underwater environments. Two tasks are involved: 
\begin{itemize}
    \item \textbf{ShipsEar:}
    % This task identifies the vessel or sound-source type from underwater acoustic recordings. It distinguishes 12 categories, including different types of fishing, service, recreational, and commercial vessels, as well as natural ambient noise.
    This task identifies the vessel or sound-source type from underwater acoustic recordings. It distinguishes 12 categories: Fish boat, Trawler, Mussel boat, Tugboat, Dredger, Motorboat, Pilot boat, Sailboat, Passenger ship, Ocean liner, RORO ship, and Natural ambient noise. 
    \item \textbf{DeepShip:}
    This task identifies the type of vessel from underwater acoustic recordings. It distinguishes four categories: Cargo, Passengership, Tanker, and Tug. 
\end{itemize}

\textbf{Bioacoustics.} Bioacoustics studies biological sounds and their use in understanding and identifying living organisms. One  task is involved: 
\begin{itemize}
    \item \textbf{MarmAudio:}
    This task identifies the vocalization type of marmoset calls. It distinguishes six categories: Twitter, Tsik, Phee, Trill, Infant cry, and Seep.
    % \item \textbf{BEANS-HumBugDB:}
    % This task performs mosquito recognition based on wingbeat sounds. It distinguishes 14 categories, covering multiple mosquito species and species groups, as well as non-mosquito and other sounds.
\end{itemize}

\textbf{Manufacturing.}
Manufacturing involves monitoring industrial equipment and diagnosing machine conditions to ensure reliable operation. Two tasks are involved: 
\begin{itemize}
    \item \textbf{SDUST-bear:}
    % This task performs bearing fault diagnosis based on vibration signals. It distinguishes 10 operating conditions, including the normal condition and multiple inner-race, outer-race, and rolling-element fault settings.
    This task performs bearing fault diagnosis based on vibration signals. It distinguishes 10 categories: NC, IF0.2, IF0.4, IF0.6, OF0.2, OF0.4, OF0.6, RF0.2, RF0.4, and RF0.6. Here, NC denotes the normal condition, while IF, OF, and RF denote inner-race, outer-race, and rolling-element faults, respectively.
    \item \textbf{PU-cur:}
    This task performs bearing fault diagnosis based on motor current signals. It distinguishes three categories: healthy, inner-race (IR) fault, and outer-race (OR) fault.
    % \item \textbf{IIEE:}
    % This task identifies the operating condition of an electric engine from recorded signals. It distinguishes three categories: good, broken, and heavy load.
\end{itemize}

\subsection{Dataset Description}
\label{apdx:dataset_detailed}

This section provides a description of the datasets contained in MHB.

\textbf{GWOSC}~\citep{wu2026scits} is a synthetic gravitational-wave dataset designed for gravitational-wave event detection. It is constructed by injecting gravitational-wave templates generated through numerical relativity simulations into real detector strain data obtained from the Gravitational Wave Open Science Center (GWOSC), thereby preserving realistic detector noise. The noise pool consists of 4-second segments sampled at 16~kHz and extracted from high-quality detector intervals. Waveform templates spanning typical binary black hole mass ranges are scaled to different matched-filter signal-to-noise ratios and superimposed on these noise segments. 

\textbf{LEAVES}~\citep{fei2024leaves} is a large-scale light-curve dataset designed for the automatic classification of variable stars. It is constructed by merging open data from three major astronomical surveys: the All-Sky Automated Survey for SuperNovae (ASAS-SN), Gaia, and the Zwicky Transient Facility (ZTF). The dataset comprises 977,953 light curves from variable stars and 134,592 from nonvariable sources. 
% All variable stars are annotated with seven superclasses: Eclipsing binaries, Rotational variables, RR Lyrae, Cepheids, Long-period variables, Delta Scuti, and Nonvariable.

% \textbf{StarEmbed}~\citep{li2025starembed} is a multi-band light-curve dataset designed for the classification of variable stars. It contains approximately 40,000 light curves collected by the Zwicky Transient Facility (ZTF), covering seven stellar variability classes: EW, EA, RRab, RRc, RRd, RS CVn, and LPV. Each object contains photometric observations from the $g$ and $r$ bands. The two bands are sampled independently and may therefore contain different numbers of observations.

\textbf{Iquique}~\citep{woollam2019convolutional} is a seismic waveform dataset containing earthquake recordings from the aftershock sequence following the 2014 Iquique earthquake (moment magnitude $M_w = 8.1$) in northern Chile. The dataset contains 13,400 examples of picked seismic arrivals, all of which correspond to earthquake events. The recordings are collected from three-component seismic stations and sampled at 100~Hz. We obtain the dataset through SeisBench~\citep{woollam2022seisbench}. 

\textbf{Llaima}~\citep{llaimadataset2020} is a volcanic seismic dataset designed for the classification of seismic events. It contains 3,592 signals recorded at the LAV station of Llaima Volcano in Chile between 2010 and 2016. The signals correspond to the vertical component of seismic recordings sampled at 100~Hz and were processed to retain the representative segment of each event. 
% Each signal is annotated by specialists from the Observatorio Vulcanológico de los Andes Sur (OVDAS) into one of four categories: Volcano-Tectonic (VT), Long Period (LP), Tremor (TR), and Tectonic (TC).
The signals are annotated by specialists from the Observatorio Vulcanol\'ogico de los Andes Sur (OVDAS). 

% \textbf{Yasur}~\citep{toney2022yasur} is a labeled volcano infrasound dataset designed for subcrater classification of explosive events. It contains more than 7,500 explosions recorded by a five-station infrasound network at Yasur Volcano, Vanuatu. The explosions are localized using backprojection and associated with one of the volcano's two summit subcraters. Each event is therefore labeled according to its subcrater of origin, forming a binary classification dataset.

\textbf{Yaseen}~\citep{yaseen2018classification} is a phonocardiogram (PCG) dataset designed for heart sound classification. It contains 1,000 recordings collected from books and online sources, with all signals standardized to a sampling rate of 8~kHz. 
% The dataset is evenly divided into five categories with 200 recordings per class: Normal, Aortic Stenosis, Mitral Regurgitation, Mitral Stenosis, and Mitral Valve Prolapse. These categories cover normal heart sounds and four common valvular heart diseases.
The dataset is evenly divided into five categories, with 200 recordings per class, covering normal heart sounds and four common valvular heart diseases. 

% \textbf{CirCor}~\citep{oliveira2021circor} is a large-scale pediatric phonocardiogram (PCG) dataset containing heart sound recordings collected from multiple auscultation locations. It comprises over 5,000 recordings from 1,568 subjects, with recordings acquired at the main cardiac auscultation sites, including the aortic, pulmonary, tricuspid, and mitral valve areas. The original dataset provides detailed annotations for heart murmurs and auscultation locations. In MHB, we use only the auscultation location information and formulate a four-class classification task over the aortic, tricuspid, pulmonary, and mitral locations.
\textbf{CirCor}~\citep{oliveira2021circor} is a large-scale pediatric phonocardiogram (PCG) dataset containing over 5,000 heart sound recordings from 1,568 subjects, acquired at multiple cardiac auscultation sites. The original dataset provides detailed annotations for heart murmurs and auscultation locations. In MHB, we use only the auscultation-location labels to define \textbf{CirCor-loc}, a four-class classification task. 

% \textbf{PhysioNet FPCG}~\citep{cesarelli2012simulation} is a fetal phonocardiogram (fPCG) dataset containing real fetal heart sound recordings from pregnant women during the last trimester of pregnancy. It comprises 26 recordings collected from healthy singleton pregnancies with gestational ages ranging from 31 to 40 weeks. The recordings were acquired using a portable phonocardiographic device and sampled at 333~Hz, with an average recording duration of approximately 20 minutes. Gestational age is provided for each recording and is used as the prediction target in MHB.

\textbf{RadSeg (Radar Segmentation Dataset)}~\citep{huang2024radseg} is a synthetic radar dataset originally designed for sample-wise segmentation of interleaved radar pulse activities. It contains five radar signal types with signal-to-noise ratios ranging from $-20$ to $20$~dB at a resolution of $0.5$~dB, where each sequence consists of 32,768 complex baseband I/Q samples acquired at 3.2~MHz. In MHB, we use the sample-wise annotations to extract signal segments containing a single radar class, thereby reformulating the original segmentation task as radar segment classification.

% \textbf{RadioML 2016.10A}~\citep{o2016radio} is a synthetic radio signal dataset designed for automatic modulation classification. RadioML 2016.10A contains 220,000 I/Q signal samples spanning 11 modulation types, including eight digital modulations and three analog modulations. Each sample consists of 128 complex baseband I/Q observations, and the signal-to-noise ratio ranges from $-20$ to $18$~dB in increments of $2$~dB. The 11 modulation classes are 8PSK, AM-DSB, AM-SSB, BPSK, CPFSK, GFSK, PAM4, QAM16, QAM64, QPSK, and WBFM.
\textbf{RadioML 2016.10A}~\citep{o2016radio} is a synthetic radio signal dataset designed for automatic modulation classification. It contains 220,000 I/Q sequences spanning 11 modulation types, including eight digital modulations and three analog modulations. Each sequence consists of 128 complex baseband I/Q observations, and the signal-to-noise ratio ranges from $-20$ to $18$~dB in increments of $2$~dB.

% \textbf{ShipsEar}~\citep{santos2016shipsear} is a real-world underwater acoustic dataset designed for vessel sound recognition, containing recordings of 11 vessel types together with natural ambient noise. We follow the 12-class train--test split introduced by Xie et al.~\citep{xie2024adversarial}\footnote{\url{https://github.com/xy980523/ShipsEar-An-Unofficial-Train-Test-Split}}, which partitions the original recordings at the file level to reduce overlap between training and test data. 
\textbf{ShipsEar}~\citep{santos2016shipsear} is a real-world underwater acoustic dataset designed for vessel sound recognition, containing recordings of 11 vessel types together with natural ambient noise. We follow the 12-class train--test split introduced by \citet{xie2024adversarial}\footnote{\url{https://github.com/xy980523/ShipsEar-An-Unofficial-Train-Test-Split}}, which partitions the original recordings primarily at the file level to reduce overlap between training and test data. Following this split and the corresponding segmentation protocol, we obtain 625 signal segments, including 541 training samples and 84 test samples, covering all 12 sound-source categories.

% \textbf{DeepShip}~\citep{irfan2021deepship} is a large-scale real-world underwater acoustic dataset designed for vessel classification. It contains 47~hours and 4~minutes of recordings from 265 vessels across four classes: Cargo, Passengership, Tanker, and Tug, collected under diverse sea states and noise conditions. We follow a publicly released unofficial train--test split\footnote{\url{https://github.com/ZhuPengsen/Method-for-Splitting-the-DeepShip-Dataset}}, which partitions the data at the recording level to avoid overlap between training and test segments. Following this split and our segmentation procedure, we obtain 56,255 signal segments, including 39,371 training samples and 16,884 test samples.
\textbf{DeepShip}~\citep{irfan2021deepship} is a large-scale real-world underwater acoustic dataset designed for vessel classification. It contains 47~hours and 4~minutes of recordings from 265 vessels across four vessel classes, collected under diverse sea states and noise conditions. We follow a publicly released unofficial train--test split\footnote{\url{https://github.com/ZhuPengsen/Method-for-Splitting-the-DeepShip-Dataset}}, which partitions the data at the recording level to avoid overlap between training and test segments. Following this split and our segmentation procedure, we obtain 56,255 signal segments, including 39,371 training samples and 16,884 test samples, from which 527 test instances are included in MHB.

\textbf{MarmAudio}~\citep{lamothe_2025_15017207} is a large-scale dataset of common marmoset vocalizations designed for call-type classification. It contains more than 800,000 audio segments totaling approximately 253 hours, recorded at a sampling rate of 96~kHz. 
% Among them, approximately 215,000 calls are annotated with six vocalization types: Infant cry, Phee, Seep, Trill, Tsik, and Twitter.
Among them, approximately 215,000 calls are labeled with one of six vocalization types.

% \textbf{BEANS}~\citep{hagiwara2023beans,kiskin2021humbugdb} is a mosquito wingbeat sound dataset used for species classification. The original HumBugDB contains large-scale acoustic recordings of mosquitoes collected across diverse environments and includes annotations for multiple mosquito species. We use the HumBugDB classification subset provided in BEANS, where species with 100 or fewer samples are grouped into an \texttt{OTHER} category, resulting in a 14-class classification task.

\textbf{SDUST}~\citep{han2022novel} is a rolling-bearing vibration dataset collected on a dedicated fault-diagnosis test bench at Shandong University of Science and Technology. It contains vibration signals recorded under normal operation and multiple inner-race, outer-race, and rolling-element fault conditions, with a sampling rate of 25.6~kHz. 
% In MHB, we use the processed SDUST bearing subset released with FISHER/RMIS~\citep{fan2026fisher}, which defines a 10-class fault-diagnosis task consisting of the normal condition (NC) and inner-race (IF), outer-race (OF), and rolling-element (RF) faults at three fault sizes: 0.2, 0.4, and 0.6.
In MHB, we use the processed SDUST bearing subset released with FISHER/RMIS~\citep{fan2026fisher} for \textbf{SDUST-bear}, a 10-class bearing fault-diagnosis task.

\textbf{PU}~\citep{lessmeier2016bearing} is a bearing condition-monitoring dataset from Paderborn University containing synchronously recorded motor current and vibration signals under diverse bearing conditions. The original dataset comprises measurements from 32 bearing experiments, including healthy bearings and bearings with artificially induced or naturally developed damage, recorded under four operating conditions. 
% In MHB, we use the motor-current subset processed in FISHER/RMIS~\citep{fan2026fisher} and formulate a three-class bearing fault diagnosis task consisting of healthy, inner-race fault (IR), and outer-race fault (OR).
In MHB, we use the motor-current subset processed in FISHER/RMIS~\citep{fan2026fisher} for \textbf{PU-cur}, a three-class bearing fault-diagnosis task.

% \textbf{IIEE}~\citep{grollmisch2019sounding} is an industrial acoustic dataset designed for electric-engine condition monitoring. It contains 2,378 mono audio recordings sampled at 44.1~kHz and covers three operating states: good, broken, and heavy load. The recordings include both clean conditions and several types of background interference, such as speech, white noise, factory ambience, and stress-test perturbations. Following the original protocol, only the clean recordings are used for training, while recordings with additional background noise are reserved for evaluation, resulting in a substantially larger test set.

\section{EVALUATED MODELS}
\label{apdx:model_access}

This section provides corresponding model families, citations, and access points.

\subsection{Audio LLMs}

\textbf{Qwen2.5-Omni}~\citep{xu2025qwen25omni} is an end-to-end multimodal model supporting text, image, audio, and video inputs, with text and speech generation capabilities. We evaluate the 3B and 7B variants on MHB using the checkpoints \texttt{Qwen/Qwen2.5-Omni-3B} and \texttt{Qwen/Qwen2.5-Omni-7B}. \emph{Access:} \url{https://huggingface.co/Qwen/Qwen2.5-Omni-3B} and \url{https://huggingface.co/Qwen/Qwen2.5-Omni-7B}.

\textbf{Audio Flamingo 3}~\citep{ghosh2025audioflamingo3} is a fully open 7B large audio-language model designed for audio understanding and reasoning across speech, sound, and music. We evaluate Audio Flamingo 3 on MHB using the official checkpoint \texttt{nvidia/audio-flamingo-3}. \emph{Access:} \url{https://github.com/NVIDIA/audio-flamingo} and \url{https://huggingface.co/nvidia/audio-flamingo-3}.

\textbf{Gemini 3 Flash}~\citep{google2026gemini3flash} is a multimodal model from Google supporting text, image, video, audio, and PDF inputs with text generation. We evaluate Gemini 3 Flash on MHB through the Gemini API using the model identifier \texttt{gemini-3-flash-preview}. \emph{Access:} \url{https://ai.google.dev/gemini-api/docs/models/gemini-3-flash-preview}.

\textbf{GPT Audio Mini}~\citep{openai2026gptaudio} is a cost-efficient audio-capable model from OpenAI supporting both text and audio inputs and outputs. We evaluate GPT Audio Mini on MHB through the OpenAI API using the model identifier \texttt{gpt-audio-mini}. \emph{Access:} \url{https://developers.openai.com/api/docs/models/gpt-audio-mini}.

\textbf{Doubao-Seed-2.0}~\citep{bytedance2026doubao2} is a family of general-purpose multimodal models from ByteDance, offering variants with different performance, latency, and cost trade-offs. We evaluate the Mini and Lite variants on MHB through the Volcano Engine API, using the model identifiers \texttt{doubao-seed-2-0-mini-260215} and \texttt{doubao-seed-2-0-lite-260215}, respectively. \emph{Access:} \url{https://www.volcengine.com/product/doubao/} and \url{https://www.volcengine.com/docs/82379/1795150}.

\subsection{Domain-Specific Foundation Models}

\textbf{Time-MoE}~\citep{shi2025timemoe} is a time-series foundation model designed for general-purpose forecasting across diverse temporal data. We evaluate the 200M-parameter variant on MHB using the official checkpoint \texttt{Maple728/TimeMoE-200M}. \emph{Access:} \url{https://huggingface.co/Maple728/TimeMoE-200M}.

\textbf{SpectrumFM}~\citep{zhou2025spectrumfm} is a spectrum foundation model designed for intelligent spectrum management and pretrained with self-supervised objectives including masked reconstruction and next-slot signal prediction. We evaluate SpectrumFM on MHB using the official pretrained checkpoint released for RML2018. \emph{Access:} \url{https://github.com/ChunyuLiu188/SpectrumFM}.

\textbf{SeisLM}~\citep{liu2024seislm} is a foundation model for seismic waveforms pretrained on the large-scale SeisBench waveform collection. We evaluate the official pretrained SeisLM base model on MHB. \emph{Access:} \url{https://github.com/liutianlin0121/seisLM}.

\textbf{FISHER}~\citep{fan2026fisher} is a foundation model for unified representation of heterogeneous industrial signals, including sound, vibration, and voltage signals. We evaluate the 21.4M-parameter small variant on MHB using the official checkpoint \texttt{jiangab/FISHER-small-0723}. \emph{Access:} \url{https://huggingface.co/jiangab/FISHER-small-0723}.

\textbf{SPEAR}~\citep{yang2025spear} is a self-supervised foundation model designed to learn unified representations across speech and general audio. We evaluate the 327M-parameter Large speech--audio variant on MHB using the official checkpoint \texttt{marcoyang/spear-large-speech-audio}. \emph{Access:} \url{https://huggingface.co/marcoyang/spear-large-speech-audio}.

\section{Implementation Details of MachEar}

\subsection{MachEar Architecture and Token Fusion}
\label{apdx:MachEar}

Given an input signal $x$, MachEar and the pretrained Whisper encoder process it in parallel. MachEar consists of two parallel pathways: an FFT-based spectral pathway and a dilated 1D CNN temporal pathway. 
% Both pathways are aligned to $Q$ token positions, with $Q=64$ in all experiments. Their representations, $\mathbf{F}_{\mathrm{FFT}}$ and $\mathbf{F}_{\mathrm{CNN}}$, are concatenated along the feature dimension, projected to a shared hidden space, and processed by Transformer blocks to produce a compact sequence of MachEar tokens:
Both pathways are aligned to $Q$ token positions, producing representations $\mathbf{F}_{\mathrm{FFT}}$ and $\mathbf{F}_{\mathrm{CNN}}$ with feature dimensions $D_{\mathrm{FFT}}$ and $D_{\mathrm{CNN}}$, respectively. These representations are concatenated along the feature dimension, projected into a shared hidden space of dimension $D$, and processed by $N$ Transformer blocks, each with $H$ attention heads, followed by an output projection to produce $Q$ compact MachEar tokens:

\begin{equation}
\mathbf{H}_{\mathrm{M}}
=
\mathrm{Proj}_{\mathrm{out}}
\left(
\mathrm{Transformer}
\left(
\mathrm{Proj}_{\mathrm{in}}
\left[
\mathbf{F}_{\mathrm{FFT}};
\mathbf{F}_{\mathrm{CNN}}
\right]
\right)
\right).
\label{eq:machear}
\end{equation}
For the FFT pathway, when $L \ge Q$, a signal of length $L$ is divided into $Q$ non-overlapping patches with patch length $P=\lfloor L/Q \rfloor$, and an FFT representation is computed for each patch. 
The 1D CNN instead operates directly on the full waveform, and its temporal features are adaptively pooled to the same $Q$ token positions. For inputs shorter than $Q$, a global representation is used and invalid trailing positions are masked.

In parallel, the original Whisper encoder produces audio tokens $\mathbf{H}_{\mathrm{W}}$, while the textual prompt is embedded as $\mathbf{H}_{\mathrm{T}}$. The three sequences are concatenated and jointly fed into the LLM:

\begin{equation}
\mathbf{H}_{\mathrm{input}}
=
\mathrm{Concat}_{\mathrm{seq}}
\left(
\mathbf{H}_{\mathrm{M}},
\mathbf{H}_{\mathrm{W}},
\mathbf{H}_{\mathrm{T}}
\right).
\label{eq:framework}
\end{equation}
In all MachEar configurations, we set $Q=64$, $N=6$, $D=1024$, $H=8$, and $D_{\mathrm{FFT}}=D_{\mathrm{CNN}}=512$, with seven layers in the 1D CNN. 
The same MachEar architecture and tokenization scheme are used for all machine-hearing tasks without domain-specific modifications.

\subsection{Preprocessing and Finetuning Details}
\label{apdx:setup}

\paragraph{Input preprocessing.}
Multi-channel input signals are flattened into a one-dimensional sequence before encoding. All input signals are truncated to at most 240,000 samples. 
% To better exploit the pretrained representations of Whisper, signals sampled between 2 and 96 kHz are resampled to 16 kHz before entering the Whisper pathway, while those outside this range retain their original sampling rates to avoid excessive changes in sequence length. In contrast, MachEar processes the truncated waveform at its original sampling rate.
To better align with Whisper's pretraining sampling rate of 16~kHz, signals sampled between 2 and 96~kHz are resampled to 16~kHz before entering the Whisper pathway, while those outside this range retain their original sampling rates to avoid excessive changes in sequence length. In contrast, MachEar directly processes the truncated waveform without resampling, providing complementary representations across the full range of input sampling rates.

\paragraph{Pathway adaptation and integration.}
% We consider four configurations corresponding to Fig.~\ref{fig:mh_ft}. \texttt{Whisper\_frz} uses the pretrained Whisper encoder while keeping it frozen, whereas \texttt{Whisper} jointly finetunes the Whisper encoder with the LLM. \texttt{MachEar} directly replaces the pretrained Whisper encoder with MachEar, while \texttt{Whisper + MachEar} adds MachEar alongside the pretrained audio pathway.
We consider four main configurations corresponding to Fig.~\ref{fig:mh_ft}. \texttt{Whisper\_frz} and \texttt{Whisper} retain only the original Whisper encoder of Qwen2.5-Omni-7B, with the encoder frozen and finetuned, respectively. \texttt{MachEar} replaces the original Whisper encoder with MachEar, while \texttt{Whisper + MachEar} combines both encoders and jointly finetunes them. For the component ablations in Fig.~\ref{fig:MachEar-analysis}(b), we additionally consider \texttt{Whisper* + MachEar}, which retains both encoders but keeps the pretrained Whisper encoder frozen.

\section{Response Validity and Failure Analysis}
\label{apdx:failure_modes}

% \begin{table}[t]
% \centering
% \caption{Failure decomposition for Audio LLMs. SR is the fraction of examples with
% a valid, parseable response.}
% \label{tab:failure_analysis}
% \begin{tabular}{lccc}
% \toprule
% \textbf{Model} & \textbf{Invalid} & \textbf{Wrong valid} & \textbf{SR (\%)} \\
% \midrule
% Qwen2.5-Omni-3B & -- & -- & -- \\
% Qwen2.5-Omni-7B & -- & -- & -- \\
% Audio Flamingo 3 (7B) & -- & -- & -- \\
% GPT Audio Mini & -- & -- & -- \\
% Gemini 3 Flash & -- & -- & -- \\
% Doubao 2.0 Mini & -- & -- & -- \\
% Doubao 2.0 Lite & -- & -- & -- \\
% \bottomrule
% \end{tabular}
% \end{table}

% We distinguish invalid responses (refusal, missing label, out-of-set label, or format violation) from incorrect labels among valid responses. Table~\ref{tab:failure_analysis} will report the corresponding counts together with the SR values summarized in Table~\ref{tab:baseline_full}; representative outputs and annotation rules are provided with the final counts.

In this section, we present a detailed task-level breakdown of success rates and representative invalid responses.

\paragraph{Task-level success rates.}
% Table~\ref{tab:task_level_sr} reports the success rate of each Audio LLM on the 13 MHB tasks considered in our adaptation experiments. 
Table~\ref{tab:task_level_sr} reports the success rate of each Audio LLM across the 13 MHB tasks. 
A response is considered valid only when it conforms to the predefined task-specific label set and output format. 
% The overall success rates reported in Fig.~\ref{fig:sr&fc} are computed over all 6,128 samples rather than by averaging task-level rates. 
The overall success rates reported in Fig.~\ref{fig:sr&fc} are computed over all 6,128 instances rather than by averaging task-level rates. 
Task-level success rates vary substantially across input regimes. In particular, LEAVES, RadSeg, and RadioML exhibit frequent input-handling failures for several models, whereas many tasks within more conventional audio regimes achieve near-perfect response validity.

\begin{table*}[t]
\centering
\caption{
% Task-level success rates (\%) of Audio LLMs on the 13 MHB tasks used in our adaptation experiments. A response is considered successful only if it matches a predefined task-specific label and satisfies the required output format.
Task-level success rates (\%) of Audio LLMs on the 13 MHB tasks. A response is considered successful only if it matches the predefined task-specific label set and required output format.
}
\label{tab:task_level_sr}
\vspace{-0.8em}
\setlength{\tabcolsep}{3.5pt}
\renewcommand{\arraystretch}{1.06}

\resizebox{\textwidth}{!}{%
\begin{tabular}{@{}llccccccc@{}}

% =========================================================
% Upper part
% =========================================================
\toprule

\multirow{2}{*}{\makecell[l]{\textbf{Access}}}
&
\multirow{2}{*}{\textbf{Model}}
&
\multicolumn{2}{c}{\textbf{Astronomy}}
&
\multicolumn{2}{c}{\textbf{Earth Science}}
&
\multicolumn{3}{c}{\textbf{Healthcare}}
\\

\cmidrule(lr){3-4}
\cmidrule(lr){5-6}
\cmidrule(lr){7-9}

& &
\textbf{GWOSC}
&
\textbf{LEAVES}
&
\textbf{Iquique}
&
\textbf{Llaima}
&
\textbf{Yaseen}
&
\multicolumn{2}{c}{\textbf{CirCor-loc}}
\\

\midrule

\multirow{3}{*}{\textbf{Open-source}}
&
Qwen2.5-Omni-3B
& 100.00
& 7.13
& 100.00
& 100.00
& 100.00
& \multicolumn{2}{c}{80.92}
\\

&
Qwen2.5-Omni-7B
& 100.00
& 7.13
& 73.71
& 100.00
& 91.50
& \multicolumn{2}{c}{83.79}
\\

&
Audio-Flamingo-3-7B
& 100.00
& 100.00
& 100.00
& 100.00
& 100.00
& \multicolumn{2}{c}{100.00}
\\

\midrule

\multirow{4}{*}{\textbf{Proprietary}}
&
Gemini-3-Flash
& 100.00
& 5.47
& 100.00
& 96.37
& 100.00
& \multicolumn{2}{c}{100.00}
\\

&
GPT-Audio-Mini
& 4.88
& 0.00
& 40.24
& 33.52
& 100.00
& \multicolumn{2}{c}{90.82}
\\

&
Doubao-2.0-Mini
& 99.80
& 0.39
& 100.00
& 100.00
& 100.00
& \multicolumn{2}{c}{95.98}
\\

&
Doubao-2.0-Lite
& 100.00
& 0.39
& 98.80
& 100.00
& 100.00
& \multicolumn{2}{c}{99.86}
\\

\midrule
\midrule

% =========================================================
% Lower part
% =========================================================

\multirow{3}{*}{\makecell[l]{\textbf{Access}}}
&
\multirow{3}{*}{\textbf{Model}}
&
\multicolumn{2}{c}{\textbf{Electromagnetic}}
&
\multicolumn{2}{c}{\textbf{Underwater Acoustics}}
&
\multicolumn{1}{c}{\textbf{Bioacoustics}}
&
\multicolumn{2}{c}{\textbf{Manufacturing}}
\\

\cmidrule(lr){3-4}
\cmidrule(lr){5-6}
\cmidrule(lr){7-7}
\cmidrule(lr){8-9}

& &
\textbf{RadSeg}
&
\textbf{RadioML}
&
\textbf{ShipsEar}
&
\textbf{DeepShip}
&
\makecell{\textbf{Marm}\\\textbf{Audio}}
&
\makecell{\textbf{SDUST-}\\\textbf{bear}}
&
\makecell{\textbf{PU-}\\\textbf{cur}}
\\

\midrule

\multirow{3}{*}{\textbf{Open-source}}
&
Qwen2.5-Omni-3B
& 52.70
& 0.00
& 100.00
& 100.00
& 99.67
& 100.00
& 100.00
\\

&
Qwen2.5-Omni-7B
& 52.07
& 0.00
& 100.00
& 100.00
& 99.33
& 100.00
& 100.00
\\

&
Audio-Flamingo-3-7B
& 100.00
& 100.00
& 100.00
& 0.00
& 100.00
& 100.00
& 100.00
\\

\midrule

\multirow{4}{*}{\textbf{Proprietary}}
&
Gemini-3-Flash
& 86.93
& 80.00
& 100.00
& 100.00
& 94.33
& 100.00
& 100.00
\\

&
GPT-Audio-Mini
& 0.00
& 0.00
& 86.90
& 15.75
& 1.67
& 75.70
& 41.88
\\

&
Doubao-2.0-Mini
& 0.00
& 0.00
& 100.00
& 99.81
& 100.00
& 100.00
& 100.00
\\

&
Doubao-2.0-Lite
& 0.00
& 0.00
& 100.00
& 100.00
& 100.00
& 100.00
& 100.00
\\

\bottomrule

\end{tabular}%
}

% \vspace{-0.6em}
\end{table*}

\paragraph{Failure categories and representative cases.}
We categorize invalid responses into four types. 
% \textbf{Unrecognized Input} denotes cases where the model behaves as if the provided signal were absent and requests the input again. 
\textbf{Unrecognized Input} denotes cases where the model appears not to recognize that a signal has been provided, treats it as missing, and requests the input again.
\textbf{Short Input} denotes failures caused by minimum-duration or encoder-length constraints. \textbf{Empty Response} denotes cases with no usable textual output. \textbf{Label/Format Error} denotes responses that do not match the predefined label set or required output format. These categories affect response validity only; 
% Macro-F1 is still computed over all MHB samples, with invalid responses treated as incorrect predictions.
Macro-F1 is still computed over all MHB instances, with invalid responses treated as incorrect predictions.

Representative examples of each failure type are shown in Table~\ref{tab:failure_cases}. Outputs are reproduced verbatim or shortened only for presentation, without changing the failure semantics.

% \begin{table*}[t]
% \centering
% \caption{
% Representative examples of the four invalid-response categories.
% Each example is selected from a different model and task.
% }
% \label{tab:failure_cases}
% \vspace{-0.8em}
% \footnotesize
% \setlength{\tabcolsep}{3pt}
% \renewcommand{\arraystretch}{1.10}

% \begin{tabularx}{\textwidth}{
% @{}
% >{\raggedright\arraybackslash}p{0.15\textwidth}
% >{\raggedright\arraybackslash}p{0.18\textwidth}
% >{\raggedright\arraybackslash}p{0.08\textwidth}
% >{\raggedright\arraybackslash}p{0.29\textwidth}
% >{\raggedright\arraybackslash}p{0.24\textwidth}
% @{}
% }
% \toprule
% \textbf{Failure Type}
% &
% \textbf{Model}
% &
% \textbf{Task}
% &
% \textbf{Representative Output / Error}
% &
% \textbf{Why Invalid}
% \\
% \midrule

% \textbf{Unrecognized Input}
% &
% Gemini-3-Flash
% &
% RadioML
% &
% ``Please provide the in-phase (I) and quadrature (Q) data components ...''
% &
% Requests the I/Q data again although the signal has already been provided.
% \\

% \addlinespace[0.35em]

% \textbf{Short Input}
% &
% Qwen2.5-Omni-3B
% &
% LEAVES
% &
% ``Calculated output size ... 0. Output size is too small.''
% &
% The input is too short for the audio encoder.
% \\

% \addlinespace[0.35em]

% \textbf{Empty Response}
% &
% Doubao-2.0-Mini
% &
% GWOSC
% &
% \textit{[empty output]}
% &
% No usable textual response is returned.
% \\

% \addlinespace[0.35em]

% \textbf{Label/Format Error}
% &
% Audio-Flamingo-3-7B
% &
% DeepShip
% &
% GT: \texttt{Tanker};\quad
% Response: \texttt{Tank}
% &
% \texttt{Tank} does not exactly match the predefined label \texttt{Tanker}.
% \\

% \bottomrule
% \end{tabularx}
% \end{table*}

\begin{table}[t]
\centering
\caption{Representative examples of the four invalid-response categories. Each example is selected from a different model and task.}
\label{tab:failure_cases}
\vspace{-0.8em}
\renewcommand{\arraystretch}{1.15}
\begin{tabularx}{\linewidth}{@{}>{\raggedright\arraybackslash}p{0.18\linewidth}@{\hspace{1.5em}}>{\raggedright\arraybackslash}X@{}}
\toprule
\textbf{Failure Type} & \textbf{Representative Case} \\
\midrule

\textbf{Unrecognized Input} &
\textbf{Model:} Gemini-3-Flash \quad \textbf{Task:} RadioML\par
\textbf{Output:} ``The input is missing. Please provide the in-phase and quadrature components.''\par
\textbf{Why invalid:} Treats the provided signal as missing and requests data again. \\
\midrule

\textbf{Short Input} &
\textbf{Model:} Doubao-2.0-Mini \quad \textbf{Task:} LEAVES\par
\textbf{Error:} ``The request failed because the duration of the input audio (4ms) is shorter than the limit (50ms).''\par
\textbf{Why invalid:} The input is shorter than the API's minimum duration of 50\,ms. \\
\midrule

\textbf{Empty Response} &
\textbf{Model:} Qwen2.5-Omni-3B \quad \textbf{Task:} CirCor-loc\par
\textbf{Output:} \emph{[empty output]}\par
\textbf{Why invalid:} No usable textual response is returned. \\
\midrule

\textbf{Label/Format Error} &
\textbf{Model:} Audio-Flamingo-3-7B \quad \textbf{Task:} DeepShip\par
\textbf{Output:} \texttt{Tank} \quad \textbf{Ground truth:} \texttt{Tanker}\par
\textbf{Why invalid:} \texttt{Tank} is not among the predefined labels. \\
\bottomrule
\end{tabularx}
\end{table}

% \section{Analysis by Signal Characteristics}
% \label{app:signal_characteristics}

% \section{Full Results on Our Benchmark}
% \label{app:full_results}

% \begin{table*}[t]
% \centering
% \caption{Per-domain baseline macro-F1 (\%). Avg. and rank are summarized in
% Table~\ref{tab:baseline_full}.}
% \label{tab:baseline_full}
% \resizebox{\textwidth}{!}{%
% \begin{tabular}{lccccccc}
% \toprule
% \textbf{Model} & \textbf{Astronomy} & \textbf{Geology} & \textbf{Industrial} &
% \textbf{Healthcare} & \textbf{EM} & \textbf{Bioacoustics} & \textbf{Underwater} \\
% \midrule
% Qwen2.5-Omni-3B & 12.70 & 13.30 & 7.65 & 20.26 & 3.21 & 5.35 & 7.57 \\
% Qwen2.5-Omni-7B & 12.70 & 16.42 & 9.67 & 19.95 & 3.85 & 9.66 & 7.71 \\
% Audio Flamingo 3 (7B) & 16.76 & 21.86 & 12.85 & 15.58 & 3.91 & 6.09 & 0.11 \\
% GPT Audio Mini & 2.27 & 13.20 & 11.89 & 18.51 & 0.00 & 0.00 & 5.73 \\
% Gemini 3 Flash & 11.87 & 28.54 & 17.22 & 15.76 & 8.46 & 14.52 & 15.25 \\
% Doubao 2.0 Mini & 10.91 & 24.31 & 7.41 & 10.98 & 0.00 & 17.85 & 13.25 \\
% Doubao 2.0 Lite & 11.03 & 25.25 & 5.46 & 18.46 & 0.00 & 19.51 & 15.72 \\
% \midrule
% Time-MoE & 64.44 & 66.51 & 27.56 & 42.41 & 42.66 & 15.61 & 19.13 \\
% SpectrumFM & 58.75 & 61.60 & 28.40 & 37.56 & 62.28 & 22.45 & 29.57 \\
% SeisLM & 55.20 & 74.12 & 47.32 & 43.38 & 64.89 & 29.11 & 45.87 \\
% FISHER & 22.92 & 0.00 & 66.67 & 50.97 & 0.00 & 92.57 & 45.18 \\
% SPEAR & 56.18 & 70.28 & 57.98 & 60.62 & 48.02 & 77.55 & 54.19 \\
% \bottomrule
% \end{tabular}%
% }
% \end{table*}

\section{Case examples}
\label{apdx:case}

In this section, we present representative case examples from MHB. 
% For visualization, the corresponding machine-hearing signals are plotted alongside the task instructions and model responses.
Each example includes a visualization of the input signal, the task instruction, and the corresponding model response.

\vspace{1.0em}

\begin{adjustbox}{scale=0.95,center}
\begin{tcolorbox}[
    width=\linewidth,
    colback=white,          % 内容背景色
    colframe=green!20!black,% 边框颜色
    coltitle=white,         % 标题文字颜色
    colbacktitle=green!20!black, % 标题栏背景色
    % title=Astronomy: gravitational wave events,
    title={Astronomy: GWOSC, gravitational-wave event detection},
    fonttitle=\bfseries,
    boxrule=0.8mm,          % 边框粗细
    arc=2mm,                % 圆角
    left=2mm, right=2mm, top=1mm, bottom=1mm
]
\begin{center}
\includegraphics[height=0.18\textheight, keepaspectratio]{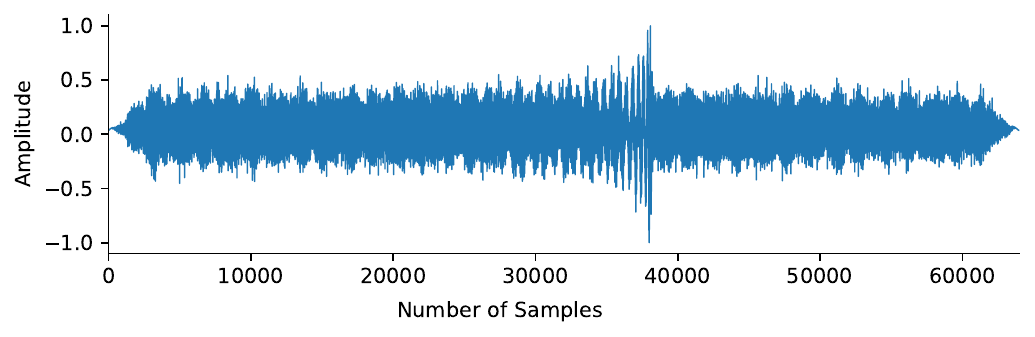}\\
\end{center}
\textbf{Question:}\\
Please analyze the provided time-series data and determine whether a gravitational wave event has occurred. Classify the result as ``normal'' (no gravitational wave event detected) or ``anomaly'' (a gravitational wave event is detected).\\
\textbf{Answer:}\\
The classification result is: anomaly.
\end{tcolorbox}
\end{adjustbox}

\vspace{1.0em}

\begin{adjustbox}{scale=0.95,center}
\begin{tcolorbox}[
    width=\linewidth,
    colback=white,          % 内容背景色
    colframe=green!20!black,% 边框颜色
    coltitle=white,         % 标题文字颜色
    colbacktitle=green!20!black, % 标题栏背景色
    % title=Earth Science: earthquake source depth,
    title={Earth Science: Iquique, earthquake focal-depth classification},
    fonttitle=\bfseries,
    boxrule=0.8mm,          % 边框粗细
    arc=2mm,                % 圆角
    left=2mm, right=2mm, top=1mm, bottom=1mm
]
\begin{center}
\includegraphics[height=0.22\textheight, keepaspectratio]{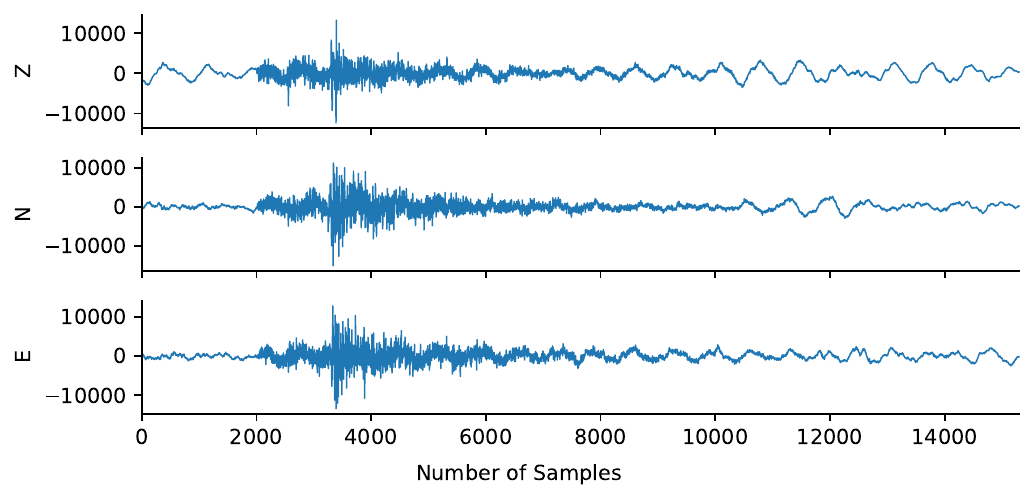}\\
\end{center}
\textbf{Question:}\\
Now I give you a focal depth classification task. Classify the focal depth of the earthquake source into one of the following categories: shallow (0-70 km), intermediate (70-300 km), or deep (300-700 km). Analyze the recording and determine which category it belongs to.\\
\textbf{Answer:}\\
The focal depth of this seismic event is classified as shallow.
\end{tcolorbox}
\end{adjustbox}

\begin{adjustbox}{scale=0.95,center}
\begin{tcolorbox}[
    width=\linewidth,
    colback=white,          % 内容背景色
    colframe=green!20!black,% 边框颜色
    coltitle=white,         % 标题文字颜色
    colbacktitle=green!20!black, % 标题栏背景色
    % title=Healthcare: auscultation location from a PCG recording,
    title={Healthcare: CirCor-loc, auscultation-location identification},
    fonttitle=\bfseries,
    boxrule=0.8mm,          % 边框粗细
    arc=2mm,                % 圆角
    left=2mm, right=2mm, top=1mm, bottom=1mm
]
\begin{center}
\includegraphics[height=0.17\textheight, keepaspectratio]{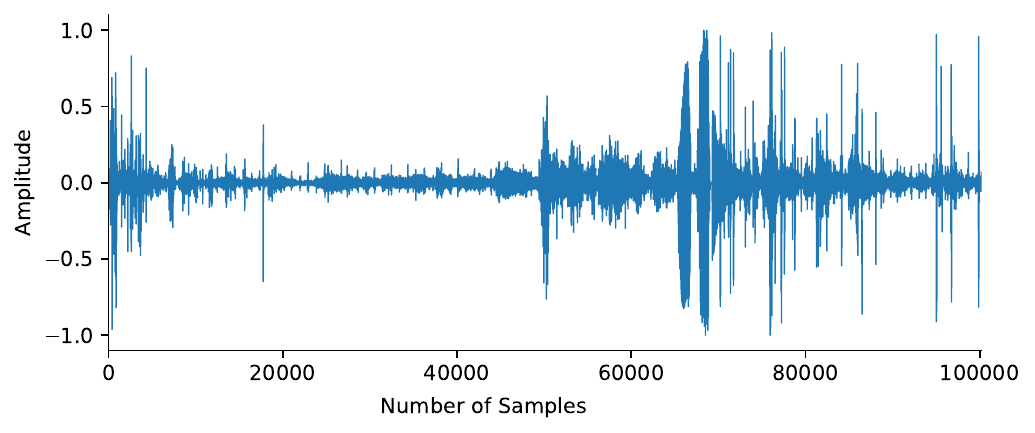}\\
\end{center}
\textbf{Question:}\\
This is a phonocardiogram (PCG) recording of heart sounds, with a single channel, a sampling rate of 4000 Hz, and a duration of 25.06 seconds. Based on this PCG signal, please identify the auscultation location (aortic valve, tricuspid valve, pulmonary valve, or mitral valve).\\
\textbf{Answer:}\\
This recording is identified as pulmonary valve.
\end{tcolorbox}
\end{adjustbox}

\begin{adjustbox}{scale=0.95,center}
\begin{tcolorbox}[
    width=\linewidth,
    colback=white,          % 内容背景色
    colframe=green!20!black,% 边框颜色
    coltitle=white,         % 标题文字颜色
    colbacktitle=green!20!black, % 标题栏背景色
    % title=Electromagnetic: radio signal modulation type,
    title={Electromagnetic: RadioML, radio modulation classification},
    fonttitle=\bfseries,
    boxrule=0.8mm,          % 边框粗细
    arc=2mm,                % 圆角
    left=2mm, right=2mm, top=1mm, bottom=1mm
]
\begin{center}
\includegraphics[height=0.17\textheight, keepaspectratio]{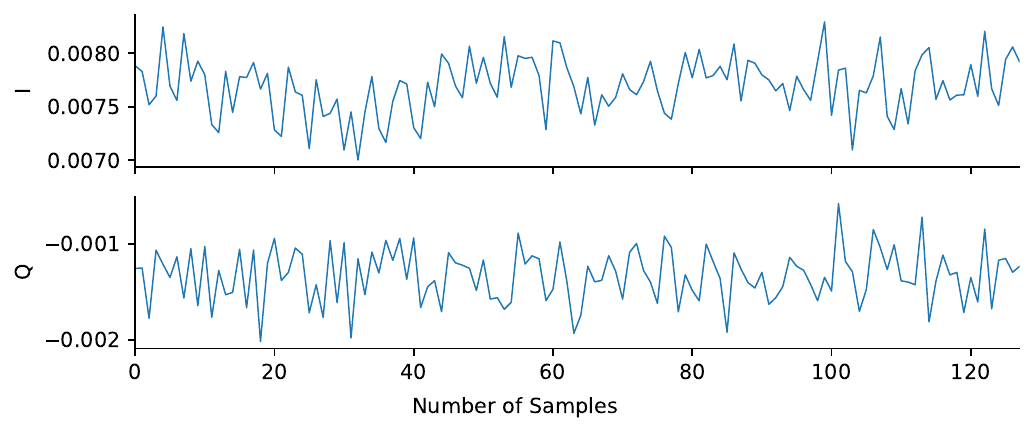}\\
\end{center}
\textbf{Question:}\\
Given this two-channel I/Q sequence, identify the modulation scheme. The answer must be one of: 8PSK, AM-DSB, AM-SSB, BPSK, CPFSK, GFSK, PAM4, QAM16, QAM64, QPSK, WBFM.\\
\textbf{Answer:}\\
The modulation type is AM-SSB.
\end{tcolorbox}
\end{adjustbox}

\begin{adjustbox}{scale=0.95,center}
\begin{tcolorbox}[
    width=\linewidth,
    colback=white,          % 内容背景色
    colframe=green!20!black,% 边框颜色
    coltitle=white,         % 标题文字颜色
    colbacktitle=green!20!black, % 标题栏背景色
    % title=Underwater Acoustics: vessel type,
    title={Underwater Acoustics: DeepShip, vessel-type classification},
    fonttitle=\bfseries,
    boxrule=0.8mm,          % 边框粗细
    arc=2mm,                % 圆角
    left=2mm, right=2mm, top=1mm, bottom=1mm
]
\begin{center}
\includegraphics[height=0.17\textheight, keepaspectratio]{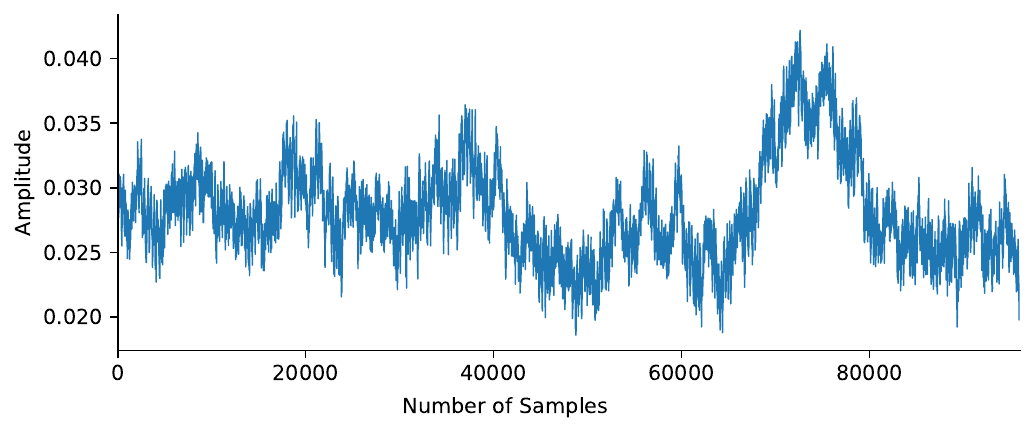}\\
\end{center}
\textbf{Question:}\\
Given an underwater acoustic recording from the DeepShip dataset, which contains real-world underwater recordings of commercial vessels collected in the Strait of Georgia, classify the vessel into one of the following 4 categories: Cargo, Passengership, Tanker, Tug. Analyze the recording and determine which category it belongs to.\\
\textbf{Answer:}\\
The recording is classified as Tanker.
\end{tcolorbox}
\end{adjustbox}

\begin{adjustbox}{scale=0.95,center}
\begin{tcolorbox}[
    width=\linewidth,
    colback=white,          % 内容背景色
    colframe=green!20!black,% 边框颜色
    coltitle=white,         % 标题文字颜色
    colbacktitle=green!20!black, % 标题栏背景色
    % title=Bioacoustics: vocalization type of marmoset calls,
    title={Bioacoustics: MarmAudio, marmoset vocalization-type classification},
    fonttitle=\bfseries,
    boxrule=0.8mm,          % 边框粗细
    arc=2mm,                % 圆角
    left=2mm, right=2mm, top=1mm, bottom=1mm
]
\begin{center}
\includegraphics[height=0.22\textheight, keepaspectratio]{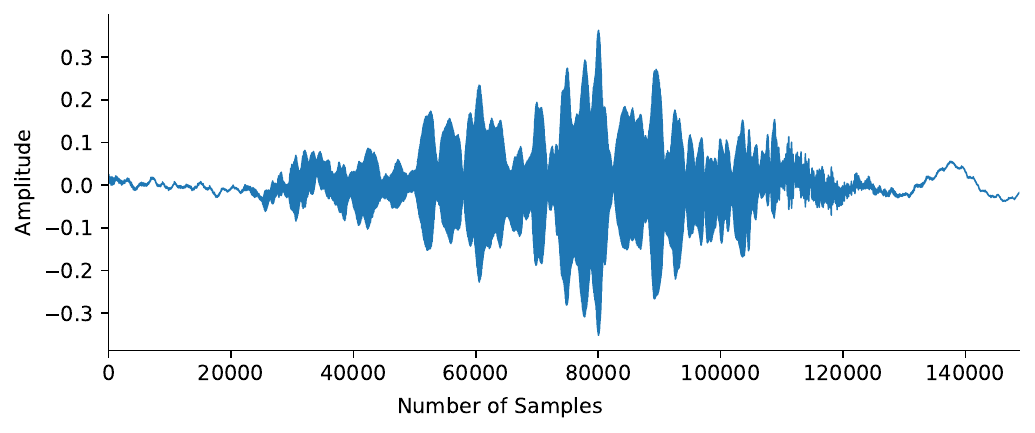}\\
\end{center}
\textbf{Question:}\\
Now, I need your assistance to complete a marmoset classification task. The provided audio is a recording of a marmoset call. It is known that the provided data corresponds to one of the following categories: Twitter, Tsik, Phee, Trill, Infant\_cry or Seep. Please determine which of the following call types it belongs to.\\
\textbf{Answer:}\\
This audio is classified as Phee.
\end{tcolorbox}
\end{adjustbox}

\vspace{1.0em}

\begin{adjustbox}{scale=0.95,center}
\begin{tcolorbox}[
    width=\linewidth,
    colback=white,          % 内容背景色
    colframe=green!20!black,% 边框颜色
    coltitle=white,         % 标题文字颜色
    colbacktitle=green!20!black, % 标题栏背景色
    % title=Manufacturing: bearing fault diagnosis from vibration,
    title={Manufacturing: SDUST-bear, vibration-based bearing fault diagnosis},
    fonttitle=\bfseries,
    boxrule=0.8mm,          % 边框粗细
    arc=2mm,                % 圆角
    left=2mm, right=2mm, top=1mm, bottom=1mm
]
\begin{center}
\includegraphics[height=0.22\textheight, keepaspectratio]{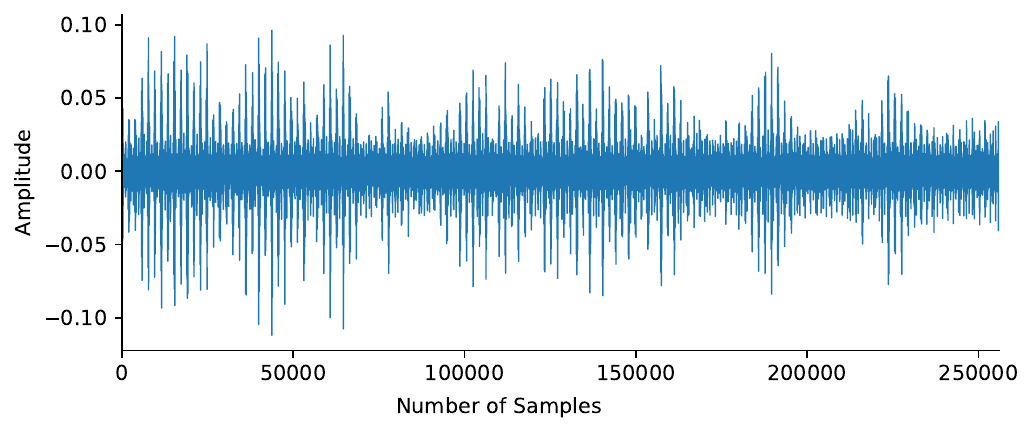}\\
\end{center}
\textbf{Question:}\\
Now, I need your assistance to complete a fault classification task. This is a bearing fault diagnosis task. It is known that the provided vibration signal sample corresponds to one of the following categories: NC, IF0.2, IF0.4, IF0.6, OF0.2, OF0.4, OF0.6, RF0.2, RF0.4, RF0.6. Please identify the classification result represented by this data.\\
\textbf{Answer:}\\
The classification result is: RF0.4.
\end{tcolorbox}
\end{adjustbox}

\end{document}

%% file: math_commands.tex
\usepackage{amsmath,amsfonts,bm}

\def\eqref#1{equation~\ref{#1}}
\def\1{\bm{1}}

\DeclareMathAlphabet{\mathsfit}{\encodingdefault}{\sfdefault}{m}{sl}
\SetMathAlphabet{\mathsfit}{bold}{\encodingdefault}{\sfdefault}{bx}{n}